\documentclass[%
 reprint,
 amsmath,amssymb,
 aps,
floatfix,
]{revtex4-2}

\usepackage{graphicx}
\usepackage{dcolumn}
\usepackage{bm}
\usepackage{hyperref}

\usepackage{physics} 
\usepackage{float}
\begin{document}

\preprint{APS/123-QED}

\title{Reducing the instability of an optical lattice clock using multiple atomic ensembles}

\author{Xin Zheng$^{1}$}
\author{Jonathan Dolde$^{1}$}%
\author{Shimon Kolkowitz$^{1,2}$}
\email{kolkowitz@berkeley.edu}

\affiliation{%
 $^{1}$Department of Physics, University of California, Berkeley, CA 94720, USA
}%

\affiliation{%
 $^{2}$Department of Physics, University of Wisconsin-Madison, Madison, WI 53706 USA
}%

\date{\today}

\begin{abstract}
The stability of an optical atomic clock is a critical figure of merit for almost all clock applications. To this end, much optical atomic clock research has focused on reducing clock instability by increasing the atom number, lengthening the coherent interrogation times, and introducing entanglement to push beyond the standard quantum limit.
In this work, we experimentally demonstrate an alternative approach to reducing clock instability using a phase estimation approach based on individually controlled atomic ensembles in a strontium (Sr) optical lattice clock.
We first demonstrate joint Ramsey interrogation of two spatially-resolved atom ensembles that are out of phase with respect to each other, which we call ``quadrature Ramsey spectroscopy,'' resulting in a factor of 1.36(5) reduction in absolute clock instability as measured with interleaved self-comparisons.
We then leverage the rich hyperfine structure of ${}^{87}$Sr to realize independent coherent control over multiple ensembles with only global laser addressing.
Finally, we utilize this independent control over four atom ensembles to implement a form of phase estimation, achieving a factor of greater than $3$ enhancement in coherent interrogation time and a factor of 2.08(6) reduction in instability over an otherwise identical single ensemble clock with the same local oscillator and the same number of atoms. We expect that multi-ensemble protocols similar to those demonstrated here will result in reduction in the instability of any optical lattice clock with an interrogation time limited by the local oscillator.
\end{abstract}

\maketitle


\section{Introduction}

Thanks to their remarkable precision and accuracy, optical atomic clocks are rapidly advancing the frontiers of timekeeping, quantum science, and fundamental physics~\cite{ludlow_optical_2015}.
State-of-the-art optical clocks have now reached the level of $10^{-18}$ in both stability and accuracy~\cite{SingleIon_-18_2016,mcgrew_atomic_2018,bothwell_jila_2019,Brewer_IonClock_2019,Oelker_ClockStability_2019},
enabling novel emerging applications such as relativistic geodesy, searches for ultralight dark matter, and gravitational wave detection~\cite{chou_optical_2010,derevianko_hunting_2014,takano_geopotential_2016,kolkowitz_gravitational_2016,PTB_transportable_2017,grotti_geodesy_2018,safronova_search_2018,takamoto_test_2020,kennedy_dark_2020,bothwell_resolving_2022}.
Instability is a critical figure of merit for optical atomic clocks, determining the sensitivity of clock comparisons to target signals in fundamental physics applications and the achievable precision in applications to timekeeping, metrology, and navigation~\cite{ludlow_optical_2015,safronova_search_2018}.  
The fundamental limit to the instability of an optical clock with unentangled atoms is given by the quantum projection noise (QPN),
\begin{equation}
    \sigma_{\rm QPN}(\tau) = \frac{1}{2\pi \nu CT}\sqrt{\frac{T + T_{\rm d}}{N \tau}},
\label{QPN}
\end{equation}
where $\nu$ is the clock transition frequency,
$C$ is the contrast,
$T$ is the coherent interrogation time,
$T_{\rm d}$ is the dead time,
$\tau$ is the averaging time,
and $N$ is the number of atoms being interrogated.

Equation~\ref{QPN} implies that the clock instability can be reduced through longer coherent interrogation times. However, in optical clocks both conventional Ramsey and Rabi spectroscopy are limited by frequency noise of the local oscillator (LO), or clock laser, used to interrogate the atoms. In the case of Ramsey spectroscopy, the interrogation time is limited by the requirement that the phase accumulated by the atoms in the rotating frame must remain within $[-\pi/2, \pi/2]$ in order to avoid phase slips. In addition, for atomic clock operations with duty cycles less than 1, noise on the LO during the dead time leads to Dick noise,
an aliased noise at harmonic frequencies of $1/T_{\rm c}$ ($T_{\rm c} = T + T_d$ corresponds to the cycle time) that compromises the clock instability~\cite{Dick1987,Santarelli1998}. 
In particular,
the Dick-limited clock instability for Ramsey spectroscopy can be approximated as (see Appendix~\ref{Appendix:DickScaling} for details)
\begin{equation}
    \sigma_{\rm Dick}(\tau) =  \frac{\sigma_{\rm LO}}{\sqrt{2{\rm ln}2}} \sum_{m=1}^{\infty}\bigg|\frac{{\rm sin}(\pi mT/T_{\rm c})}{\pi mT/T_{\rm c}}\bigg| \sqrt{\frac{T_{\rm c}}{m\tau}},
    \label{scaling_Dick}
\end{equation}
where $\sigma_{\rm LO}$ is the flicker frequency noise floor limited instability of the LO.
From Eq.~\ref{scaling_Dick},
we can see that for an atomic clock limited by Dick noise,
the instability no longer scales with atom number, but can still be reduced by increasing the interrogation time.
These factors have motivated considerable efforts into the development of optical LOs with longer coherence times~\cite{kessler_SiCavity_2012,Matei_SiCavity_2017,Zhang_SiCavity_2017,Robinson_SiCavity_2019,Kedar_SiCavity_2023}, the use of synchronous differential comparisons to push the interrogation time beyond the coherence time of the LO~\cite{takamoto_beyond_2011,schioppo_ultrastable_2017,Clements_IonCoherence_2020,zheng_differential_2022}, the use of multiple separate optical atomic clocks to actively perform feedback or feedforward on the LO~\cite{Kim_coherenceAtomicSpecies_2022}, as well as the use of entangled atomic states to push beyond the QPN limit for the same LO coherence time~\cite{Pedrozo_entanglementclock_2020,Robinson_entanglement_2022,Rydberg_entanglement_2023}, although the latter case will only be useful in scenarios where the clock is QPN-limited.

\begin{figure*}
\includegraphics[width=0.95\textwidth]{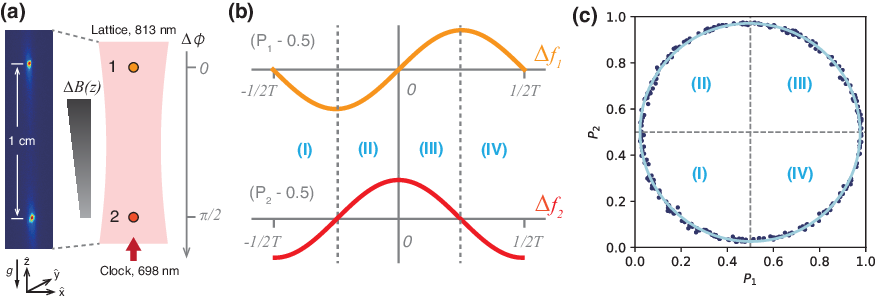}
\caption{\label{Fig1}Quadrature Ramsey spectroscopy with a multi-ensemble clock.
(a) Two ensembles of strontium atoms separated by 1 cm are prepared using a movable optical lattice.
A magnetic field gradient $\Delta B(z)$ is applied to introduce a differential phase shift ($\Delta\phi$) of $\pi/2$ between the two ensembles for a given Ramsey interrogation time $T$. 
(b) Theory curves showing the expected Ramsey fringes for the two atom ensembles with a differential phase of $\pi/2$~\cite{Li_theory_2022}. The 4 regions (I, II, III, and IV) correspond to quadratures for decoding the clock laser detuning.
(c) Parametric plot of the measured excitation fractions from the two ensembles ($P_1, P_2$) for a $\pi/2$ phase difference at a short Ramsey free evolution time ($T=7.5$ ms) for the full randomly sampled range of detunings corresponding to $\Delta f_{i}\in [-1/2T,1/2T]$, with the light blue curve a fit to the expected ellipse. Each labeled quadrature corresponds to the matching region in Fig. 1(b) (see Appendix~\ref{Appendix:PhaseComparison}).
}
\end{figure*}

A proposed alternative and complementary path to reducing the instability of optical atomic clocks is to more efficiently allocate the available atomic resources. For example, in Ref.~\cite{rosenband_ensemble_2013}, Rosenband and Leibrandt argued that if the atoms in a clock are split up into multiple ensembles and probed with different interrogation times in parallel, the clock instability can be reduced exponentially with atom number rather than with $1/\sqrt{N}$ as in Eq.~\ref{QPN} for uniform interrogation.
Related work on efficient atomic clock operations with several atomic ensembles was also proposed by Borregaard and Sørensen~\cite{borregaard_ensembles_2013}.
Similarly, in a recent theoretical work~\cite{Li_theory_2022}, Li et al.~argued that through joint interrogation of two atom ensembles that are 90 degrees out of phase,
it should be feasible to extend the interpretable phases accumulated by the atoms from $[-\pi/2, \pi/2]$ to $[-\pi, \pi]$, 
thereby doubling the achievable Ramsey interrogation time for a given LO and reducing the corresponding clock instability.

There have been several prior and recent works on combining separate optical atomic clocks to realize a single compound clock with reduced instability and longer coherent interrogation times~\cite{schioppo_ultrastable_2017,DD_2020,QND_PRX_2020,Kim_coherenceAtomicSpecies_2022}. For example, in Dörscher et al.~\cite{DD_2020}, dynamical decoupling sequences were used along with near synchronous interrogation of two separate optical atomic clocks in order to overcome the laser coherence limit, extending the coherent interrogation time by a factor of approximately 6.5.
Bowden et al.~\cite{QND_PRX_2020} performed QND measurements of one ensemble using an optical cavity to extend the interrogation time of a second ensemble in a separate apparatus by a factor of 7.
A simplified three-ensemble version of the Rosenband and Leibrandt scheme was also recently demonstrated by Kim et al.~\cite{Kim_coherenceAtomicSpecies_2022} using two separate interleaved Yb optical lattice clocks to realize zero-dead-time operation and feed-forward on the interrogation of an additional Al${}^+$ single ion clock,
achieving lifetime limited interrogation times.
Related techniques have also been implemented in atom interferometers to increase their dynamic range~\cite{interferometry_2020}.
However, the use of multiple independent optical clock apparatus, including vacuum chambers, atom sources, and associated optics, is extremely resource intensive and is not easily scaled to larger numbers of ensembles, and any compound clock would still benefit from reductions in the instability of the constituent clocks, motivating work into realizing similar gains within individual optical clocks through the use of multiple atomic ensembles.

In this work, we demonstrate a multi-ensemble strontium optical lattice clock that takes advantage of up to 4 spatially-resolved atom ensembles within a single vacuum chamber and the same optical lattice to increase the coherent interrogation time and reduce the absolute clock instability.
We first experimentally demonstrate joint interrogation on 2 out-of-phase atom ensembles \cite{Li_theory_2022}, which we call ``quadrature Ramsey spectroscopy'',
which enables the measurement of phase shifts of up to $[-\pi, \pi]$,
and reduces the clock instability by a factor of 1.36(5), as measured in a self-comparison.
We then demonstrate a novel approach which leverages the rich hyperfine structure of ${}^{87}$Sr ($I=9/2$) to enable independent coherent control of four atom ensembles in order to implement a simple form of phase estimation algorithm~\cite{Kessler_phase_2014,Pezze_phase_2020,Pezze_phase_2021}. We prepare four atom ensembles in pairs of two distinct nuclear spin states, ($\ket{{}^1S_0, m_F=5/2}$ and $\ket{{}^3P_0, m_F=3/2}$ where $m_F$ is the hyperfine sub-level),
and demonstrate independent coherent control of each ensemble pair on two different clock transitions through frequency multiplexing.
Combining the four-ensemble scheme and the quadrature Ramsey technique, 
we further improve the interrogation time by a factor of roughly 1.7,
achieving a factor of 2 reduction in instability over a single ensemble clock with the same atom number, duty cycle, and LO.

We note that our multi-ensemble protocols will introduce additional systematic frequency shifts that will require additional characterization. However, none of these shifts are new or unique to optical lattice clocks,
and we do not expect any of them to fundamentally limit accuracy above the $10^{-18}$ level (see Appendix~\ref{Appendix:Systematics} for details).
These additional systematics will likely preclude the near term adoption of multi-ensemble protocols in the most accurate state-of-the-art laboratory-based clocks.
We instead anticipate that multi-ensemble protocols such as the ones we demonstrate here will be first be adopted in applications involving transportable clocks, where stability is a more critical figure of merit and where record setting accuracy is already impractical.

\section{Experimental apparatus and quadrature Ramsey spectroscopy}

The experimental apparatus used here is similar to the multiplexed strontium optical lattice clock presented in prior works~\cite{zheng_differential_2022,zheng_redshift_2023}.
A movable, one-dimensional optical lattice is loaded with multiple atom ensembles, for example two atom ensembles at a 1 cm separation along the lattice axis ($\hat{z}$) as shown in Fig.~\ref{Fig1}(a),
followed by nuclear spin-polarization through optical pumping and in-lattice-cooling.
A bias magnetic field applied along $\hat{x}$ defines the quantization axis, 
with a typical magnitude of 5.5 G.
A tunable magnetic field gradient along $\hat{z}$ ($\Delta B(z)$), ranging from 0 to 45 mG/cm, is applied to introduce a differential frequency shift, 
 dominated by the first-order Zeeman shift,
and thus a differential phase shift for a Ramsey sequence of duration $T$ of $\Delta\phi \approx \gamma \Delta B(z) T$, where $\gamma$ is the differential gyromagnetic ratio for the addressed clock transition, between the two ensembles.
To take full advantage of the magnetic field gradient, 
we probe on the magnetically sensitive $\ket{{}^3P_0, m_F=7/2}\rightarrow\ket{{}^1S_0, m_F=5/2}$ (noted as $\ket{e, 7/2}\rightarrow\ket{g, 5/2}$) transition,
which has a magnetic field sensitivity of roughly 564 Hz/G.
Synchronous differential Ramsey spectroscopy is performed to characterize the differential phase $\Delta\phi$~\cite{zheng_differential_2022}, and the magnetic field gradient is fine-tuned so that $\Delta\phi = \pi/2$,
where the Ramsey fringes for the two ensembles are 90 degrees out-of-phase as shown in Fig.~\ref{Fig1}(b) in order to realize quadrature Ramsey spectroscopy (Appendix~\ref{Appendix:DifferentialPhaseShift}).
A typical parametric plot taken at a short free evolution time ($T=7.5$ ms) is shown in Fig.~\ref{Fig1}(c),
where synchronous Ramsey spectroscopy is performed by randomly sampling the phase of the final global $\pi/2$-pulse and each labeled quadrature corresponds to the matching region in Fig.~\ref{Fig1}(b) (see Appendix~\ref{Appendix:PhaseComparison} for details).
The differential phase has a typical long-term differential instability below $10^{-19}$ as demonstrated in prior work~\cite{zheng_differential_2022}.

\begin{figure}
\includegraphics[width=0.45\textwidth]{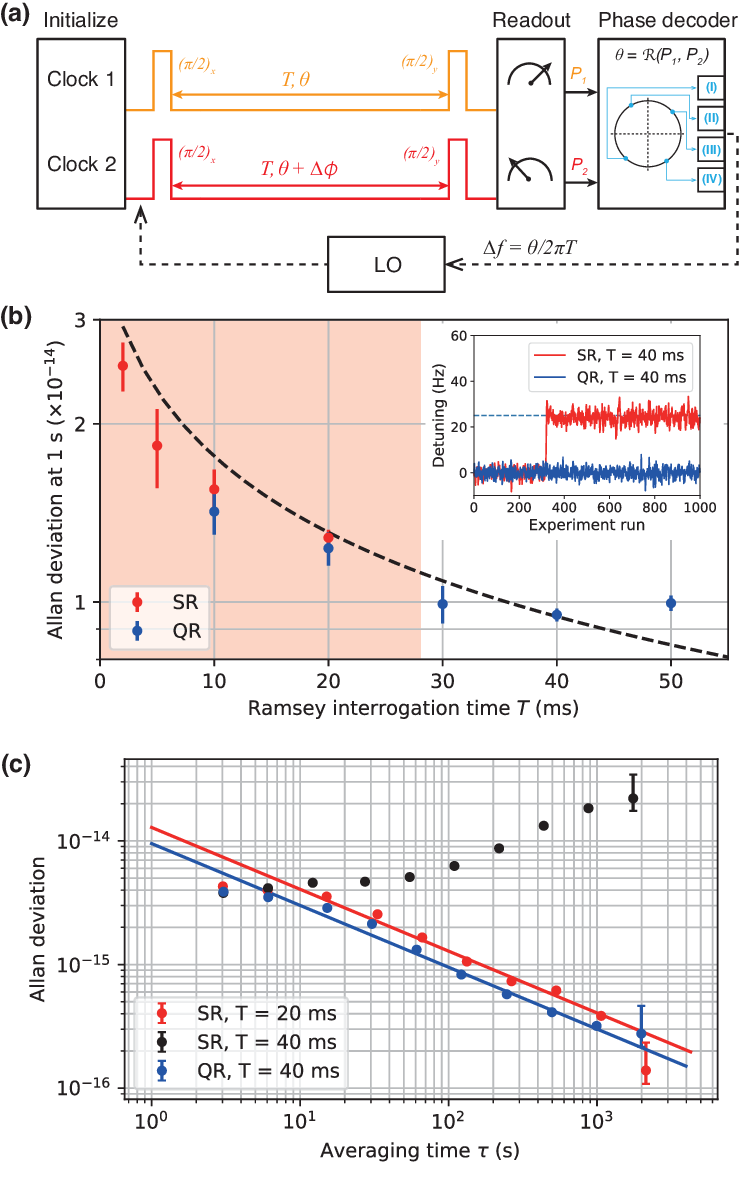}
\caption{\label{Fig2}Increasing the interrogation time and reducing clock instability with quadrature Ramsey spectroscopy.
(a) Schematic of clock operation using Ramsey spectroscopy with two atom ensembles. The differential phase, $\Delta\phi$, is tuned to either $0$ or $\pi/2$ to compare the clock performance between standard Ramsey spectroscopy and quadrature Ramsey spectroscopy.
The clock laser detuning ($\Delta f$) is decoded using the excitation fractions from both clocks,
and is then fed back to the LO.
(b) Instability as a function of Ramsey free evolution time ($T$) for both standard Ramsey (SR) with $\Delta\phi = 0$ (red) and quadrature Ramsey (QR) with $ \Delta\phi = \pi/2$ (blue).
Each point corresponds to the extracted Allan deviation of the self-comparison at 1 s, obtained through linear fit as shown in panel (c).
Error bars represent 1$\sigma$ standard deviation.
The shaded region represents a conservative estimate of the region where phase slips do not occur for SR with experiments taken over an hour for at least 2000 runs.
Dashed line is a fit to the data based on Dick noise limited instability after accounting for finite pulse length, assuming the LO noise is dominated by flicker frequency noise (see Appendix~\ref{Appendix:DickScaling}).
Inset: A representative plot of the measured frequency difference between the two interleaved servos with an interrogation time of $T=40$ ms for both SR (red) and QR (blue).
A phase slip occurs for SR, leading to a frequency offset of $1/T$ (dashed line).
(c) Lowest measured Allan deviations for clock self-comparisons for SR at $T=20$ ms (red) and QR at $T=40$ ms (blue). The Allan deviation for SR at $T=40$ ms (black), where a phase slip occurred as shown in the inset of (b), is shown for reference.
Error bars indicate 1$\sigma$ standard deviation.
The line corresponds to a fit assuming only white frequency noise ($1/\sqrt{\tau}$ scaling).
Due to the finite attack time of the servos we only fit to data after 40 s.
Each measurement averages for approximately 1 hour.
}
\end{figure}

A standard Ramsey spectroscopy sequence begins with a $\pi/2$-pulse that prepares the atoms in a 50:50 superposition of $\ket{g}$ and $\ket{e}$.
The atoms then evolve freely for time $T$ and accumulate a phase $\theta$, $\ket{g} + e^{i\theta}\ket{e}$,
in which $\theta = 2\pi T \Delta f$ and $\Delta f$ is the detuning of the clock laser from atomic resonance.
A final $\pi/2$-pulse converts $\theta$ into a population difference and generates the error signal for clock feedback (Fig.~\ref{Fig2}(a)).
However, phases beyond $[-\pi/2, \pi/2]$ result in population differences that are no longer unique, producing an erroneous error signal. The interrogation times for conventional Ramsey spectroscopy are therefore typically constrained to well below the LO coherence time in order to avoid a failure in the feedback servo due to phase slips~\cite{hume_beyond_2016}, an example of which is shown in the inset to Fig.~\ref{Fig2}(b).

Clock operation with Ramsey spectroscopy is performed on the ${}^1S_0\rightarrow{}^3P_0$ clock transition.
Because we lack a second independent optical lattice clock to perform direct comparisons against,
we instead perform a self-comparison,
where two independent atomic servos using the same experimental sequences are interleaved and compared 
in order to determine the clock instability (prior examples can be found in Refs.~\cite{selfcomp0_2011,selfcomp1_2012,selfcomp2_2015}). 
Separate digital servos correct the clock laser frequency independently for each interleaved sequence and generate a difference frequency signal which is then used to extract the Allan deviation and corresponding instability.
Ramsey spectroscopy on the $\ket{e,7/2}\rightarrow\ket{g,5/2}$ transition is performed simultaneously on two spatially resolved atom ensembles, 
each containing roughly 4000 atoms.
The interrogation time in our system is primarily limited by frequency noise of the LO,
which limits our measured atom-laser coherence time to roughly 100 ms~\cite{zheng_differential_2022}.
The typical dead time per cycle in our system is about 1.5 s, which is primarily limited by atom preparation
(roughly 1 s), and camera readout (roughly 250 ms), so there is a total of roughly 3 s of dead time for each clock in the self-comparison. 
The process of loading multiple ensembles contributes only roughly 100 ms of extra dead time (see Appendix~\ref{Appendix:Cycle} for details) and is not a limiting factor for the duty cycle in our apparatus. 

In order to perform a fair comparison between the quadrature Ramsey scheme and standard Ramsey spectroscopy using the same total atom number and experimental sequence,
the magnetic field gradient $\Delta B(z)$ is tuned such that the differential phase  between the two ensembles is either $\Delta\phi=0$ (standard Ramsey) or $\Delta\phi=\pi/2$ (quadrature Ramsey) (Fig.~\ref{Fig2}(a)).
In the case of $\Delta\phi=0$, the measurements from the two ensembles are simply averaged together and standard clock operation using Ramsey spectroscopy is performed, with the measured phase given by    
\begin{equation}
    \theta = {\rm sin}^{-1}(2(P-P_0)/C),
\label{arcsin}
\end{equation}
where $P = (P_1 + P_2)/2$ is the average of the excitation fractions from both clocks, and
$P_0\approx0.5$ and $C\approx0.95$ are the measured half-maximum and contrast of the Ramsey fringe, respectively. In this case the two ensembles are equivalent to a single ensemble clock with $2N$ total atoms.   
In the case of $\Delta\phi = \pi/2$, quadrature Ramsey spectroscopy is employed, with
the excitation fractions ($P_1, P_2$) used as inputs for the phase decoder $\mathcal{R}(P_1, P_2)$,
which determines the quadrature of the acquired phase according to~\cite{Li_theory_2022},
\begin{equation}
\begin{split}
    \theta &= \mathcal{R}(P_1, P_2)\\
         &=  \begin{cases}
      (-\pi-\theta_1 - \theta_2)/2 & \text{if $P_1<P_0$, $P_2<P_0$, (I)}\\
      (\theta_1 - \theta_2)/2 & \text{if $P_1 \le P_0$, $P_2 \ge P_0$, (II)}\\
      (\theta_1 + \theta_2)/2 & \text{if $P_1 \ge P_0$, $P_2 \ge P_0$, (III)}\\
      (\pi-\theta_1 + \theta_2)/2 & \text{if $P_1>P_0$, $P_2<P_0$, (IV)}
    \end{cases},
\end{split}
\label{decoder}
\end{equation}
in which $\theta_1 = {\rm sin}^{-1}(2(P_1-P_0)/C)$,
$\theta_2 = {\rm cos}^{-1}(2(P_2-P_0)/C)$.
We perform a series of self-comparisons for both $\Delta\phi=0$ and $\pi/2$ cases by varying the interrogation time $T$, with each measurement averaging down for roughly an hour.
Each self-comparison yields Allan deviations which are linearly fit assuming only white frequency noise (that averages down as $\tau^{-1/2}$),
which is on the order of $1\times10^{-14}/\sqrt{\tau}$ level and is dominated here by the Dick noise from the LO.
The extracted Allan deviations are plotted as a function of $T$ in Fig.~\ref{Fig2}(b).
In the case of $\Delta \phi=0$,
we find that phase slips begin to occur somewhere between the measurements at $T=20$~and $T=30$~ms, indicated by the cutoff of the shaded region in Fig.~\ref{Fig2}(b), which we conservatively place near $T=30$~ms.
This is consistent with Monte Carlo simulations of the phase slip probabilities based on the measured LO noise spectrum (Appendix~\ref{Appendix:phase_slip_probability}).
Phase slips lead to the sudden accumulation of frequency differences between the two servos at multiples of the Ramsey fringe linewidth in Fig.~\ref{Fig2}(b).
As a result, outside the shaded region in Fig.~\ref{Fig2}(b) the self-comparison no longer averages down and no meaningful Allan deviation can be defined.
In the case of quadrature Ramsey spectroscopy with $\Delta\phi=\pi/2$,
we are able to extend the interrogation time to $T=50$~ms before we begin to observe phase slips in the self-comparison.
The optimal clock performance for quadrature Ramsey spectroscopy in our apparatus is achieved at $T=40$ ms with an instability of $9.5(3)\times10^{-15}/\sqrt{\tau}$,
a factor of $1.36(5)$ smaller than the instability measured at $T=20$~ms with conventional Ramsey interrogation (Fig.~\ref{Fig2}(c)). 

\begin{figure*}
\includegraphics[width=1\textwidth]{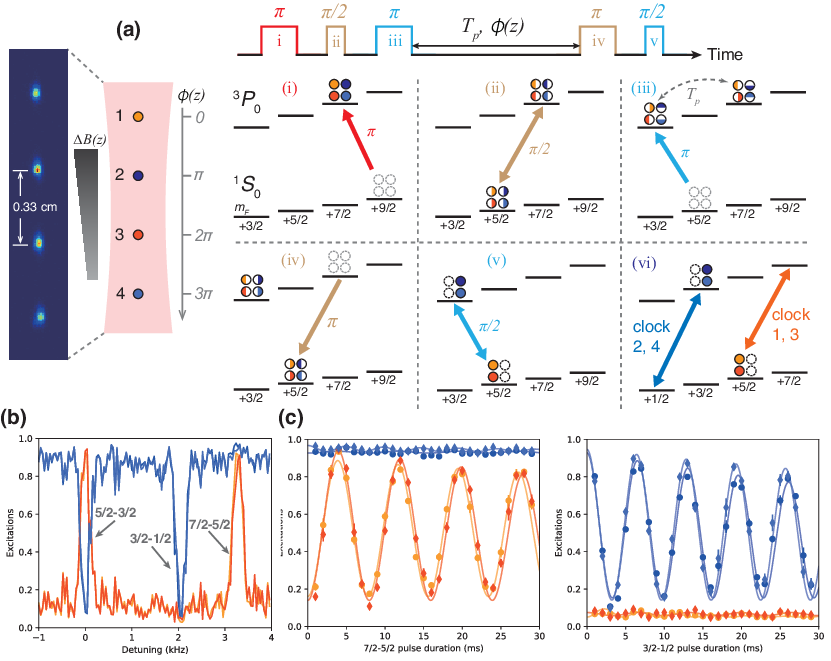}
\caption{\label{Fig3}Independent control over multiple atomic ensembles with only global addressing.
(a) Left: The lattice is loaded with 4 evenly-spaced atomic ensembles with 0.33 cm nearest-neighbor separation, and a linear magnetic field. Right: Relevant energy levels and the pulse sequence used to prepare the ensembles in different hyperfine states.
All four ensembles are initially prepared in the stretched hyperfine ground state
$\ket{{}^1S_0, m_F=9/2}$, or $\ket{g, 9/2}$.
A first $\pi$-pulse (i) transfers the population into $\ket{{}^3P_0, m_F=7/2}$ (or $\ket{e, 7/2}$), 
followed by a $\pi/2$-pulse (ii) on the $\ket{e, 7/2}\rightarrow\ket{g, 5/2}$ transition that prepares a 50:50 superposition.
A second $\pi$-pulse (iii) prepares a superposition of two nuclear spins ($\ket{e, 3/2}$ and $\ket{e, 7/2}$).
During the free evolution time $T_p$,
an applied field gradient (about $45$ G/cm) introduces a phase shift of $0$, $\pi$, $2\pi$, and $3\pi$ for each atom ensemble, respectively.
Final $\pi$-pulse (iv) and $\pi/2$-pulse (v) prepare the ensembles into opposite states, with ensembles (1, 3) on $\ket{g, 5/2}$ and ensembles (2, 4) on $\ket{e, 3/2}$.
(b) Global Rabi spectroscopy of the four atom ensembles after state preparation at a bias magnetic field of about 5.5 G.
(c) Independent coherent Rabi oscillations for ensembles (1, 3) (orange, red) on the $\ket{g, 5/2}\rightarrow\ket{e, 7/2}$ transition and for ensembles (2, 4) (blue, light blue) on the $\ket{e, 3/2}\rightarrow\ket{g, 1/2}$ transition, respectively.
Each plot represents the average of 4 independent scans, with each data point indicating the mean. Error bars correspond to 1$\sigma$ standard deviation.
The solid curves correspond to fits to the expected sinusoidal Rabi oscillations, and are used to bound the cross-talk between the two hyperfine clock transitions.
}
\end{figure*}

\section{Hyperfine state initialization and individual coherent control over ensemble pairs through global addressing}

While quadrature Ramsey spectroscopy with two ensembles enables a doubling of the coherent interrogation time, it cannot be naively extended beyond this by adding more ensembles because Ramsey fringes are inherently $2\pi$ periodic. In order to take advantage of additional ensembles and further extend the coherent interrogation time,
different interrogation times are required for different samples, as proposed in Refs.~\cite{rosenband_ensemble_2013,borregaard_ensembles_2013}, which constitute a form of a phase estimation algorithm~\cite{Kessler_phase_2014,Pezze_phase_2020,Pezze_phase_2021}. Potential approaches would be to address different spatially resolved ensembles in multiple optical lattices with different focused clock beams, or even to have different ensembles in different vacuum chambers. However, both approaches are challenging and would suffer from added differential phase~\cite{ma_phase_1994,ptb_phase_2012} and intensity noise on the clock beams, as well as additional frequency shifts due to the differing environments of each ensemble. It is therefore desirable to have all of the atomic ensembles share the same lattice, vacuum chamber, and clock beam. Fortunately, the magnetic field gradient that enables quadrature Ramsey spectroscopy can be combined with the rich hyperfine level structure of ${}^{87}$Sr ($I=9/2$) to achieve independent clock interrogation with multiple ensembles using only global addressing of all the ensembles with a single clock beam.

The basic principle and relevant pulse sequence is shown in Fig.~\ref{Fig3}(a).
Similar to the two-ensemble case,
we start with a sample of 4 evenly spaced atom ensembles, each containing roughly 2000 atoms,
in the stretched state $\ket{g, 9/2}$.
We first transfer the population into $\ket{e, 7/2}$ with a clock $\pi$-pulse (i),
then create a 50:50 superposition between $\ket{e, 7/2}$ and $\ket{g, 5/2}$ using a clock $\pi/2$-pulse (ii).
We then transfer half of the population into $\ket{e, 3/2}$ (iii),
and let the sample evolve for $T_p$,
such that each ensemble accumulates phase shifts of 0, $\pi$, $2\pi$, and $3\pi$, respectively, due to the applied linear magnetic field gradient (at about 45 mG/cm).
We then transfer the populations back to $\ket{g, 5/2}$ (iv),
and convert the accumulated phases into population in either the $\ket{g, 5/2}$ and $\ket{e, 3/2}$ states with a final clock $\pi/2$-pulse (v).
Ensembles 1 and 3 are left in $\ket{g, 5/2}$,
while ensembles 2 and 4 are left in $\ket{e, 3/2}$ (Fig.~\ref{Fig3}(a)).
This state preparation sequence typically takes about 250 ms
yielding an overall state preparation fidelity of about 90\% limited by the clock $\pi$-pulse fidelity ($>95$\% per pulse). Cleanup pulses are then applied to remove unwanted residual populations in each of ensemble pairs.

To demonstrate that the ensembles are now in two independently addressable pairs, Rabi spectroscopy is performed by scanning the frequency of the global clock laser applied to the four atom ensembles (Fig.~\ref{Fig3}(b)).
The $\pi$-pulse duration is roughly 3 ms, which power broadens the linewidth to roughly 300 Hz. The transitions remain resolved under a large bias magnetic field of 5.5 G.
Relevant transitions for the subsequent clock interrogations are the $\ket{e, 7/2}\rightarrow\ket{g, 5/2}$ (noted as $7/2-5/2$) and $\ket{e, 3/2}\rightarrow\ket{g, 1/2}$ ($3/2-1/2$) transitions,
while both ensemble pairs share the transition $\ket{g, 5/2}\rightarrow\ket{e, 3/2}$ ($5/2-3/2$).
The detuning from the magnetic field insensitive $5/2-3/2$ transition (-22.4 Hz/G) roughly indicates the ratio of the magnetic sensitivities between the two clock transitions of interest.
In particular, 
the $7/2-5/2$ transition is roughly a factor of $1.7\times$ more magnetically sensitive than the $3/2-1/2$ transition.
To characterize the cross-talk between the two clock transitions,
independent Rabi oscillations were performed on the $7/2-5/2$ ($3/2-1/2$) transition for ensemble pairs 1 and 3 (2 and 4),
as shown in Fig.~\ref{Fig3}(c).
A fit to the oscillations bounds the cross-talk between the two hyperfine clock transitions to below 3\%.

\begin{figure*}
\includegraphics[width=1\textwidth]{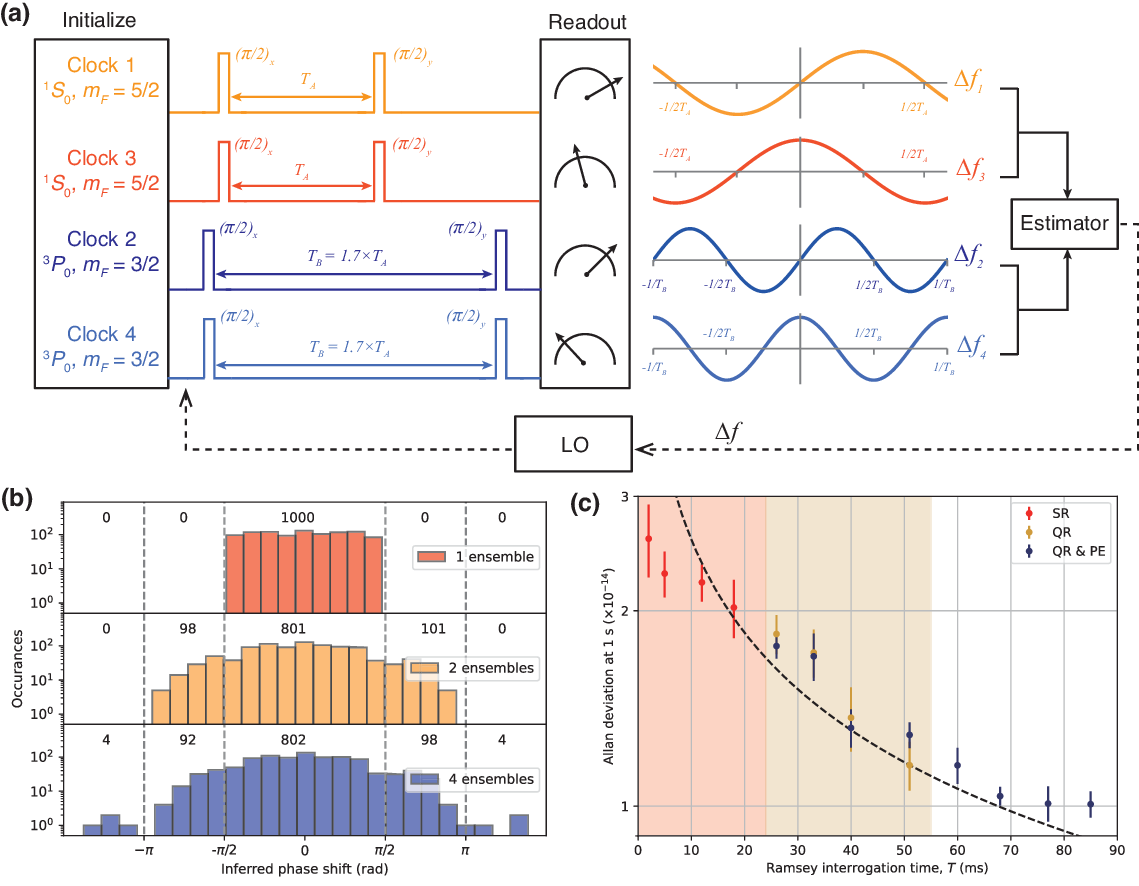}
\caption{\label{Fig4}Phase estimation using four atom ensembles with independent control through frequency multiplexing.
(a) Basic principle and pulse sequence for clock operation using phase estimation with four ensembles~\cite{rosenband_ensemble_2013,borregaard_ensembles_2013,Li_theory_2022}. We interrogate on the $\ket{e,7/2}\rightarrow\ket{g,5/2}$ transition for clock pair (1, 3) for a free evolution time of $T_A$,
during which the magnetic field gradient is chosen such that the phase difference between ensembles 1 and 3 is $\pi/2$.
Asynchronously, we interrogate on the $\ket{e,3/2}\rightarrow\ket{g,1/2}$ transition for $T_B = 1.7 \times T_A$, which is chosen such that the phase difference between ensembles 2 and 4 is also $\pi/2$ due to the difference in magnetic sensitivities for the two transitions.
An estimator extracts the LO detuning using information from all 4 clocks,
which is then used to feed back on the LO.
(b) Comparison of inferred phase shifts using various decoders for a long interrogation time. The data is taken at $(T_A, T_B)=(50, 85)$ ms in a single four-ensemble experiment for 1000 measurement cycles. The inferred phase shifts are post-processed and extracted using various decoders. Top: inferred with standard Ramsey decoder using only ensemble 2; Middle: inferred with quadrature Ramsey decoder using ensembles 2 and 4; Bottom: inferred with quadrature Ramsey with phase estimation decoder using all four ensembles.
The dashed lines indicate the $[-\pi/2,\pi/2]$ and $[-\pi, \pi]$ regions.
The numbers above the histograms represent the summed occurrences within each region. 
The y-axes are shown in log scale for clarity.
(c) Achievable interrogation time $T$ and clock instability by probing using two ensembles with $\Delta\phi=0$ (standard Ramsey, SR), two ensembles with $\Delta\phi=\pi/2$ (quadrature Ramsey, QR), and four ensembles with $\Delta\phi=\pi/2$ (quadrature Ramsey with phase estimation as shown in panel (a), QR \& PE).
Each point corresponds to the extracted Allan deviation of the self-comparison at 1 s, obtained through linear fit.
Error bars represent 1$\sigma$ standard deviation.
In the case of QR \& PE,
the interrogation time $T$ refers to $T_B$.
The shaded regions represent conservative bounds on where phase slips do not occur for each case with experiments taken over an hour for at least 2000 runs.
The dashed line is a fit to the data based on Dick noise limited instability after accounting for finite pulse length,
assuming the LO noise is dominated by flicker frequency noise (see Appendix~\ref{Appendix:DickScaling}).}
\end{figure*}

\section{Phase estimation with four atomic ensembles}
\label{sec:PE}

Having achieved independent control with global addressing through a combination of spatial and frequency multiplexing, we proceed to demonstrate a combination of quadrature Ramsey spectroscopy~\cite{Li_theory_2022} with a simple phase estimation algorithm~\cite{rosenband_ensemble_2013,borregaard_ensembles_2013} using four atomic ensembles.
This approach allows us to extend the coherent interrogation time by a factor of 1.7 over quadrature Ramsey with only two ensembles, and thereby further reduce the clock instability (see Appendix~\ref{Appendix:increasing_times}).

We interrogate a first ensemble pair to keep track of the laser phase evolution within the range of $[-\pi, \pi]$ during $T_A$ using quadrature Ramsey,
while simultaneously interrogating a second pair of ensembles on a different hyperfine clock transition with quadrature Ramsey for $T_B (T_B>T_A)$.
At the end of the measurement we perform global readout on all four ensembles, and then use a simple phase estimation algorithm to unwrap the phase accumulated by the second pair during $T_B$. As a result, we are able to unambiguously measure the phase accumulated for longer coherent interrogation times that
would otherwise lead to phase slips due to the modulo $2\pi$ ambiguity of Ramsey fringes.

The basic principle is shown in Fig.~\ref{Fig4}(a),
in which we first apply a global $\pi/2$-pulse on the $3/2-1/2$ transition for ensemble pair (2, 4),
and subsequently apply another global $\pi/2$-pulse on the $7/2-5/2$ transition for (1, 3). The delay between the asynchronous pulses for (1, 3) and (2, 4) is a few ms, limited by the rise and fall times of the clock pulses.
As the two transitions are spectrally resolved, each pulse only impacts the targeted pair, aside from a negligibly small probe shift (see Appendix~\Ref{Appendix:ProbeShift}).

We let ensembles (1, 3) freely evolve for time $T_A$,
with the magnetic field gradient $\Delta B(z)$ tuned such that the differential phase shift between ensembles (1, 3) is $\pi/2$, and a second global $\pi/2$-pulse is applied to the $7/2-5/2$ transition and the phase accumulated by the ensembles (1, 3) is stored in the populations of the clock states and is shelved there temporarily.
After a total free precession time of $T_B$ we apply the second $\pi/2$-pulse on the $3/2-1/2$ transition for ensembles (2, 4). The free precession time $T_B$ is chosen such that the differential phase of ensembles (2, 4) is again $\pi/2$,
with $T_B = 1.7\times T_A$ due to the difference in magnetic moment for the two clock transitions, resulting in two quadrature Ramsey measurements with interrogation times $T_A$ and $T_B = 1.7\times T_A$.

The populations of all four ensembles are then simultaneously read out using global imaging pulses,
where the excitation fractions are used as inputs to the phase estimator.
In particular, 
we obtain two phases from the two quadrature Ramsey measurements: $\theta_A = \mathcal{R}(P_1, P_3)$ and $\theta_B = \mathcal{R}(P_2, P_4) $.
We constrain $\theta_A$ within $[-\pi, \pi]$,
and we have $\theta_A = 2 \pi \Delta f T_A$,
in which $\Delta f$ is the LO detuning.
The actual accumulated phase shift $\theta_{{\rm act},B}$ during $T_B$ can fall outside the range of $[-\pi,\pi]$. Consequently, it cannot be handled by the two-ensemble quadrature Ramsey decoder alone (Eq.~\ref{decoder}) due to the $2\pi$ ambiguity.
To unambiguously determine $\theta_{{\rm act}, B}$,
we employ a phase estimation algorithm that uses the measurement of $\theta_A$ to remove the $2\pi$ ambiguity in $\theta_B$. This estimator is valid in the limit that the LO noise is dominated by low frequency noise and for a low clock duty cycle (see Appendix~\ref{Appendix:PhaseEstimation} for details).
We then feedback on the LO according to $\Delta f = \theta_{{\rm act}, B} /(2 \pi T_B)$.

The use of two quadrature Ramsey sequences with different interrogation times allows us to perform phase estimation and discriminate accumulated phases during $T_B$ for ensembles (2, 4) beyond $[-\pi,\pi]$. 
A representative plot from 1,000 measurement cycles of the inferred phase shifts from the longest interrogation times ($T_A=50$ ms, $T_B=85$ ms) without measured phase slips in the self-comparison using the four-ensemble approach is shown in Fig.~\ref{Fig4}(b).
We note that the inferred phase shifts are post-processed using different decoders applied to the same 1,000 measurements.
The distributions of inferred phases acquired by ensemble 2 using information from only a single ensemble (ensemble 2) with a standard Ramsey decoder (top, red), two ensembles (ensembles 2 and 4) using a quadrature Ramsey decoder (middle, yellow), and all four ensembles (ensembles 1, 2, 3, and 4) using the full phase estimation decoder (bottom, blue) are compared.
Instances of 199 inferred phases beyond $[-\pi/2, \pi/2]$ in the case of two ensembles,
and 8 inferred phases beyond $[-\pi,\pi]$ in the case of four ensembles can be observed respectively,
which would otherwise result in corresponding phase slips for the single ensemble or two-ensemble decoders for this interrogation time. 

As shown in Fig.~\ref{Fig4}(c),
the clock self-comparison Allan deviation at 1 s from all three approaches (standard Ramsey, quadrature Ramsey with two ensembles, and quadrature Ramsey combined with phase estimation with four ensembles) are compared using identical experimental sequences and atom numbers, with the only differences being the applied field gradients and how the accumulated phases are extracted by the estimator.
We are able to extend the coherent interrogation time $T_B$ out to 85 ms (with $T_A=50$ ms) without observing phase slips,
and achieve a factor of 1.2(1) reduction in instability when comparing the four-ensemble phase estimation scheme to the two-ensemble quadrature Ramsey scheme, and a factor of 2.08(6) reduction with respect to standard Ramsey. We note that the instabilities are slightly degraded over those shown in Fig.~\ref{Fig2}(b) for the same interrogation times, which we attribute to the added time required for the hyperfine state preparation (250 ms),
which contributes a total additional dead time of 500 ms in the self-comparisons.

\section{Discussion and outlook}
\label{sec:Discussion}

We note that all the instabilities and Allan deviations quoted in this work are the instabilities extracted from the self-comparisons, and that the actual single clock instabilities are expected to be at least a factor of $2$ lower due to the product of the factor of $\sqrt{2}$ from comparing two clocks and the factor of $\sqrt{2}$ from the doubling in cycle time for each clock due to the interleaved nature of the self-comparison. While the instabilities demonstrated here are not competitive with the state-of-the-art optical lattice clocks to date~\cite{mcgrew_atomic_2018,bothwell_jila_2019,Oelker_ClockStability_2019,PTB_transportable_2017,takamoto_test_2020}, this is primarily due to the relatively short coherence time of our LO ($100$~ms), which is compounded by the need for a self-comparison as we do not have another independent clock to compare with, resulting in very low duty-cycles ($<2$\%). These factors are not fundamental, and we anticipate that our multiple-ensemble techniques can be extended to other existing state-of-the-art Sr optical lattice clocks~\cite{PTB_transportable_2017,Oelker_ClockStability_2019,takamoto_test_2020} in a straightforward fashion and that similar reductions in clock instability would be expected.

For example, with state-of-the-art optical cavities that enabled atom-laser coherence time on the order of 600 ms and duty-cycle over 50\%\cite{kessler_SiCavity_2012,Matei_SiCavity_2017,Zhang_SiCavity_2017,Robinson_SiCavity_2019,Kedar_SiCavity_2023}, a factor of 2 improvement in the coherence time using quadrature Ramsey spectroscopy with two ensembles should be fairly straightforward to implement (Appendix~\ref{Appendix:HigherQualityLO}), offering roughly a factor of $\sqrt{2}$ reduction in instability.
However, given the resulting higher duty-cycle ($>50\%$), more of the frequency drift of the local oscillator will take place during the Ramsey interrogation time rather than during the dead time, so the four-ensemble phase estimation scheme employed here would likely be insufficient. We anticipate that 6 ensembles split into three quadrature Ramsey pairs, with one pair interrogated for $2T$ in parallel to two pairs each interrogated back-to-back for $T$, would be sufficient to achieve a comparable enhancement of roughly a factor of 4 in coherent interrogation time and a factor of 2 reduction in instability. However, this would require the use of an additional hyperfine clock transition or an alternative mechanism for achieving independent control over three ensemble pairs.

It is also natural to consider scaling up to many more atomic ensembles, which would eventually require an alternative approach to achieving independent control that does not require a unique clock transition for each pair of ensembles. This could be achieved in optical lattice clocks through the introduction of larger magnetic field gradients to spectroscopically resolve each ensemble on the same clock transition~\cite{boyd_nuclear_2007}, or
targeted light shifting beams to bring individual ensembles in and out of resonance with the global clock beam~\cite{xia_addressing_2015,trent_addressing_2022}.
The resulting scaling of clock instability with the number of ensembles $M$ will depend on the noise spectrum of the LO, the number of atoms available, and the clock duty cycle, 
but both the coherent interrogation time and clock instability can be expected to continue to improve with the addition of more ensembles as long as the coherent interrogation time is limited by the LO. 

While we leave the determination of the optimal algorithm and number of ensembles for a given optical atomic clock for future work, a naive extension of the 6 ensemble algorithm discussed in the previous paragraph can be expected to provide a scaling of the instability with $1/\sqrt{M}$ for large $M$, so long as the clock remains Dick noise limited. 
If combined with non-destructive mid-circuit measurement~\cite{mid_circuit_1,mid_circuit_2,mid_circuit_3,mid_circuit_4,mid_circuit_5}, the use of multiple independently controlled atom ensembles could ultimately offer exponential reductions in clock instability with atom and ensemble number using the scheme put forward by Rosenband and Leibrandt~\cite{rosenband_ensemble_2013}. Furthermore, we note that similar multi-ensemble phase estimation algorithms have been proposed to achieve Heisenber-limited clock performance using maximally entangled atomic Greenberger-Horne-Zeilinger (GHZ) states~\cite{Kessler_phase_2014}, hybrid coherent and squeezed states~\cite{Pezze_phase_2020}, and Gaussian spin-squeezed states~\cite{,Pezze_phase_2021}.

Although the multiple-ensemble protocols demonstrated in this work improve the Ramsey interrogation time and reduce the clock instability,
they do come with some trade-offs. First, in this work we incur an additional 100 ms of dead time due to multiple-ensemble preparation, although this could likely be further optimized and reduced. 
Second, our approach introduces additional systematic shifts that could impact the achievable clock accuracy. 
Examples of potential effects include Zeeman shifts arising from the magnetic field gradient applied to generate the $\pi/2$ differential phase shift for quadrature Ramsey spectroscopy,
probe shifts from the off-resonant clock pulses in the four-ensemble phase estimation approach,
and the servo error associated with the differential phase shift in the quadrature Ramsey decoder.
Estimates of the magnitude of the associated systematic shifts and potential mitigation strategies are discussed in Appendix~\ref{Appendix:Systematics}.
While a full systematic evaluation with multi-ensemble protocols is beyond the scope of this work, we estimate that the effects on achievable accuracy will be manageable at the $10^{-18}$ level. The added systematic effects therefore likely preclude the near term adoption of these techniques in optical clocks that are pushing the ultimate limits of accuracy, such as for the redefinition of the SI second. Instead, we expect that multi-ensemble techniques will initially prove most useful in situations where improved stability is beneficial and ultra-high accuracy is already either not critical or not possible, for example applications involving portable clocks such as positioning, navigation, and communication~\cite{ludlow_optical_2015,vector_atomic_2023, clock_gnss_2023,clock_navigation_2021}.  

Finally, we note that while this work has focused exclusively on protocols for extending the coherent interrogation times for Ramsey spectroscopy, many state-of-the-art optical clocks instead employ Rabi spectroscopy, and there are significant trade-offs between the two approaches. 
To the best of our knowledge,
prior works have also focused on Ramsey-like protocols to extend the interrogation times and perform phase estimation both theoretically~\cite{rosenband_ensemble_2013,borregaard_ensembles_2013,Li_theory_2022,Pezze_phase_2020,Pezze_phase_2021,Kessler_phase_2014,hume_beyond_2016} and experimentally~\cite{schioppo_ultrastable_2017,DD_2020,QND_PRX_2020,Kim_coherenceAtomicSpecies_2022}, and we are not aware of similar proposals for protocols making use of Rabi spectroscopy.
The closest example we are aware of is the combination of high-resolution imaging and Rabi spectroscopy,
as demonstrated in Marti et al.~\cite{marti_2018}, where imaging spectroscopy serves as a multiplexed measurement of laser frequency noise with a long Rabi pulse,
but in this case many of the atoms in the spatially extended ensemble contribute zero information to the measurement. We therefore leave consideration of extensions of our approaches to Rabi spectroscopy for future work. 

\section{Conclusion}

In conclusion,
in this work we experimentally demonstrated an optical lattice clock with enhanced Ramsey interrogation times and reduced absolute clock instability by harnessing multiple atomic ensembles in a single clock.
We first demonstrated quadrature Ramsey spectroscopy, a simple protocol for jointly interrogating two atom ensembles that are 90 degrees out of phase in order to extend the phase interrogation window from $[-\pi/2, \pi/2]$ to $[-\pi, \pi]$~\cite{Li_theory_2022}, extending the coherent interrogation time without phase slips by roughly a factor of $2\times$ and measuring a corresponding reduction in clock instability by a factor of 1.36(5).
We then leveraged the hyperfine structure of ${}^{87}$Sr and a linear magnetic field gradient to achieve independent state preparation of multiple ensembles with only global addressing.
Finally, we demonstrated a combination of quadrature Ramsey spectroscopy and enhanced phase estimation~\cite{rosenband_ensemble_2013,borregaard_ensembles_2013} with four ensembles 
to further extend the coherent interrogation time by a factor of 1.7 compared to the two-ensemble scheme, resulting in  a factor of 2.08(6) reduction in clock instability compared to a standard single ensemble clock with the same atom number, LO, and duty-cycle. 
Our multi-ensemble approach could be extended to other existing optical lattice clocks with state-of-the-art LOs~\cite{kessler_SiCavity_2012,Matei_SiCavity_2017,Zhang_SiCavity_2017,Robinson_SiCavity_2019,Kedar_SiCavity_2023}, and similar reductions in instability would be anticipated. In addition, we anticipate that multi-ensemble phase estimation protocols like the ones demonstrated in this work will be required to take full advantage of the metrological gains offered by entanglement such as in the use of spin-squeezed states \cite{Pedrozo_entanglementclock_2020,Robinson_entanglement_2022,Rydberg_entanglement_2023,Kessler_phase_2014,Pezze_phase_2020,Pezze_phase_2021}.

\begin{acknowledgments}
We thank Jeff Thompson, Adam Kaufman, Vladan Vuletić, and Andrew Jayich for fruitful discussions and insightful comments on the manuscript. We acknowledge technical contributions from H.-M.~Lim and N.~Ranabhat. 
This work was supported by the NIST Precision Measurement Grants program, the Northwestern University Center for Fundamental Physics and the John Templeton Foundation through a Fundamental Physics grant, the Wisconsin Alumni Research Foundation, a Packard Fellowship for Science and Engineering,  the Army Research Office through agreement number W911NF-21-1-0012, the Sloan Foundation, and the National Science Foundation under Grants No.~2143870 and 2326810.

\end{acknowledgments}

\textit{Authors' note:} While completing this work we became aware of a recent related and complementary work in Ref.~\cite{endres_multi_2023}.

\appendix

\section{Atomic sample preparation}

The experiment begins with a standard two-stage magneto-optical trap (MOT), trapping and cooling ${}^{87}$Sr atoms to a temperature of $1~\mu$K.
We then load multiple ensembles into a movable one-dimensional (1D) optical lattice as described and demonstrated in prior works~\cite{zheng_differential_2022,zheng_redshift_2023},
followed by spin-polarization and in-lattice-cooling both axially (sideband cooling) and radially (Doppler cooling) on the narrow-linewidth ${}^1S_0\rightarrow {}^3P_1$ transition at 689 nm.
The typical lattice trap depth for loading and cooling is at $70~E_{\rm rec}$,
where $E_{\rm rec}/h\approx 3.5$ kHz is the recoil energy of a lattice photon at the magic wavelength around 813.4 nm.
The lattice is then adiabatically ramped down to the operational trap depth at $15~E_{\rm rec}$ for state preparation and clock interrogation,
after which it is ramped back up to $70~E_{\rm rec}$ for read-out.

\section{Differential phase shift generation and calibration}
\label{Appendix:DifferentialPhaseShift}

\begin{figure}
\includegraphics[width=0.45\textwidth]{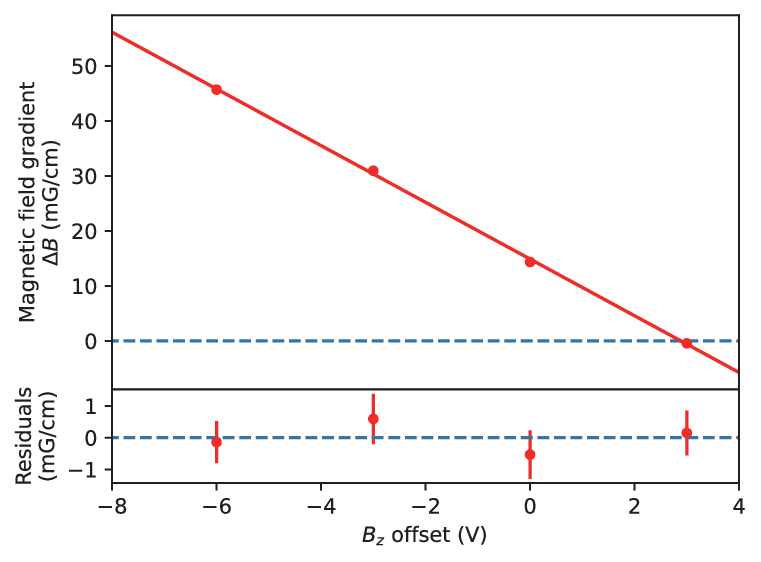}
\caption{\label{Fig5} Magnetic field gradient ($\Delta B$) adjustment by tuning the offset voltage of the $B_z$ bias coils. The B field gradient is measured on an ensemble pair separated by 1 cm via spectroscopy on the ${}^1S_0\rightarrow{}^3P_1$ transition.
Each data point has a $1\sigma$ standard deviation of roughly 0.7 mG/cm.
The solid line is a linear fit to the data.
The fit residuals are shown in the lower panel.
}
\end{figure}

A magnetic field of roughly 5.5 G is applied along $\hat{x}$ to define the quantization axis
using 3 pairs of orthogonal Helmholtz bias coils, ($B_x, B_y, B_z$).
The coil set-points are optimized to zero out $B_y$ and $B_z$ at the waist of the lattice, $z = 0$.
To introduce the $\pi/2$ differential phase shift between a pair of ensembles at $z=\pm 0.5$ cm,
we employ a magnetic field gradient which results in a differential Zeeman shift between the two ensembles. 
The magnetic field gradient can be tuned by scanning an offset voltage applied to both $B_z$ bias coils,
as shown in Fig.~\ref{Fig5}.
To take full advantage of the magnetic field gradient,
we interrogate on the $\ket{e, 7/2}\rightarrow\ket{g, 5/2}$ clock transition,
which has a magnetic field sensitivity of about $\gamma= 564$ Hz/G.
The differential frequency shift is dominated by the differential first-order Zeeman shift as,
\begin{equation}
    \Delta f_{\rm ZS, 1^{st}} = \gamma \Delta B ,
\end{equation}
where $\Delta B$ is the magnetic field gradient.
This corresponds to a typical differential frequency shift of 25 Hz for a magnetic field gradient of $\Delta B = 45$ mG/cm.


\begin{figure}[t]
\includegraphics[width=0.45\textwidth]{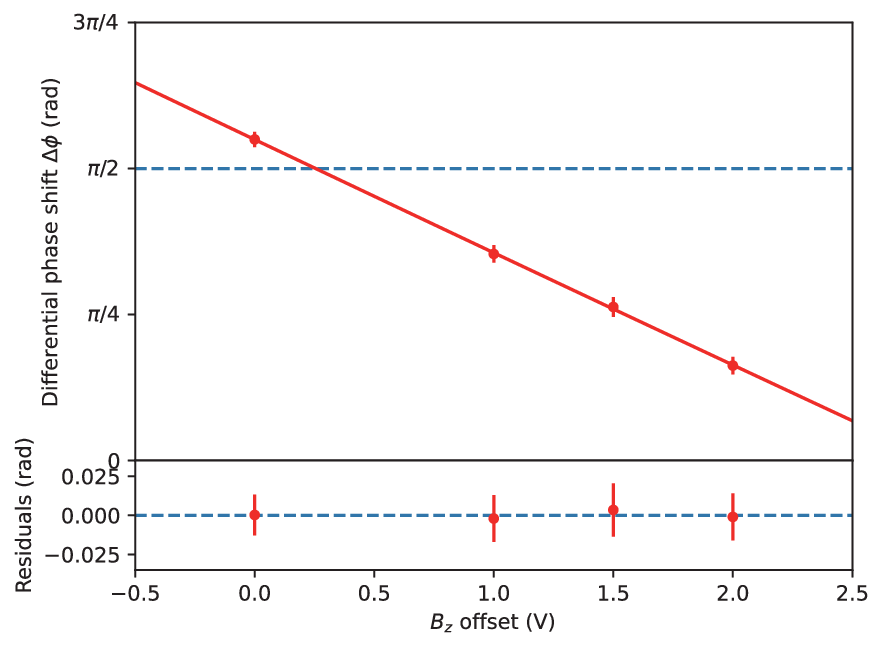}
\caption{\label{Fig6} Differential phase shift adjustment by fine tuning the $B_z$ coils offset voltage applied to both z-coils. The differential phase shift $\Delta \phi$ is measured via synchronous Ramsey spectroscopy with free evolution time of $T=30$ ms on the magnetically sensitive $\ket{e, 7/2}\rightarrow\ket{g, 5/2}$ transition.
The solid line is a quadratic fit to the data.
The residuals are shown in the lower panel.
Error bars represent $1\sigma$ standard deviation.
}
\end{figure}

The magnetic field gradient is then fine-tuned such that the differential phase $\Delta\phi = 2\pi \Delta f_{\rm ZS, 1^{st}}T$ equals to $\pi/2$ at each interrogation time $T$.
In particular, we scan the $B_z$ offset set-points and map out $\Delta \phi$ using synchronous Ramsey spectroscopy~\cite{zheng_differential_2022},
in which the two ensembles are prepared in a 50:50 superposition of ${}^1S_0$ and ${}^3P_0$,
freely evolve for $T$ and project using a final global $\pi/2$-pulse with its phase ($\phi_{\rm 2nd}$) randomly sampled from $[-\pi, \pi]$.
An ellipse fitting is then fit to the data using the least-squares method to extract $\Delta \phi$,
which has a typical uncertainty of $\sim10$ mrad, limited by the QPN.
The resulting $\Delta \phi$ as a function of $B_z$ offset voltage is fitted using a quadratic function, and the intercept of the fit with $\pi/2$ is used as the $B_z$ offset set-point for a specific Ramsey evolution time $T$,
as shown in Fig.~\ref{Fig6}.

We note that the $\partial B_z/\partial z$ gradient potentially rotates the quantization axis along the lattice direction,
which would introduce sizable tensor Stark shift gradient. 
In our previous work~\cite{zheng_redshift_2023},
we have characterized the tensor Stark shift gradient to be $-8(1)\times10^{-20}/E_{\rm rec}$/cm on the $\ket{{}^1S_0, m_F = \pm 5/2}\leftrightarrow \ket{{}^3P_0, m_F = \pm 3/2}$ transition.
This corresponds to a frequency gradient of $-1.2(2)\times10^{-18}$/cm at the operational lattice depth of $15~E_{\rm rec}$.
In future work, it appears to be feasible to mitigate this effect by generating a more uniform linear $B_x$ gradient through the addition of Golay coils~\cite{Golay_coils_1958},
requiring an upgrade to our experimental apparatus that is currently underway.

\section{Illustration of the quadrature Ramsey decoder}
\label{Appendix:PhaseComparison}

\begin{figure}[t]
\includegraphics[width=0.49\textwidth]{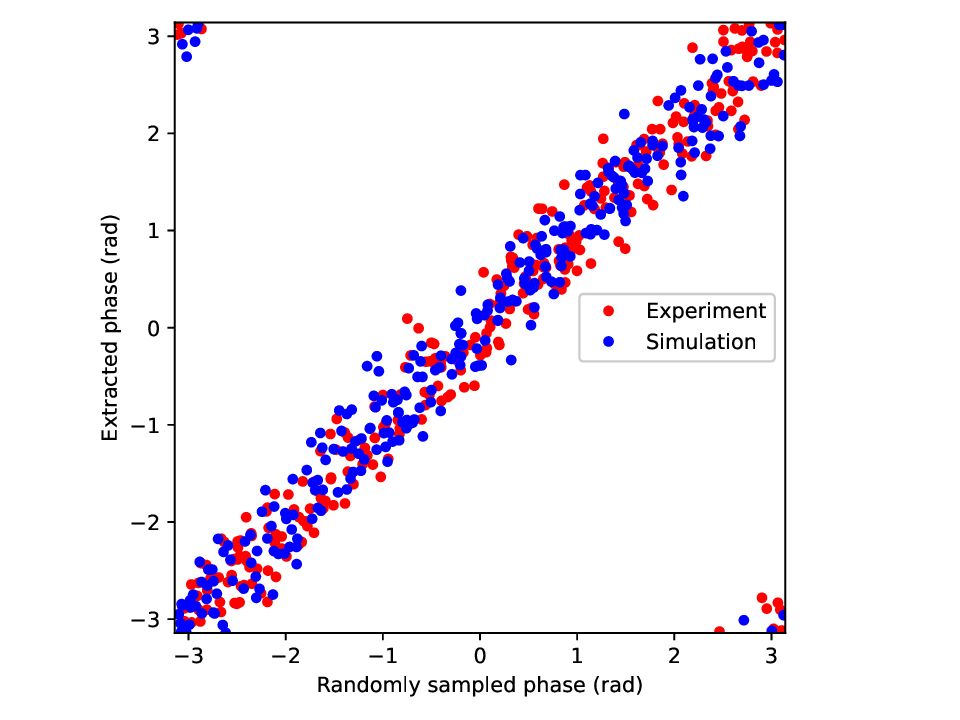}
\caption{\label{Fig7} Point-by-point comparison between known randomly applied phase shifts $\phi_{\rm 2nd} \in [-\pi,\pi]$ (to the phase of the final global $\pi/2$-pulse in synchronous Ramsey spectroscopy) and the phase shifts extracted from the decoder using quadrature Ramsey spectroscopy (Eq.~\ref{decoder}). The red scatter points correspond to the experimental data shown in Fig.~\ref{Fig1}(c) for a short Ramsey free evolution time (T = 7.5 ms). The blue scatter points correspond to Monte-Carlo simulations using the QPN and LO noise spectrum as input parameters and are shown for comparison.
}
\end{figure}

In Fig.~\ref{Fig1}(c),
we perform synchronous Ramsey spectroscopy by randomly sampling the phase ($\phi_{\rm 2nd}$) of the final $\pi/2$-pulse from a uniform distribution within $[-\pi,\pi]$.
This is equivalent to randomly sampling the LO detunings corresponding to $\Delta f_{i}\in [-1/2T,1/2T]$ under a short Ramsey free evolution time ($T=7.5$ ms),
where each quadrature can be mapped to a specific region as shown in Fig.~\ref{Fig1}(b).
To quantify the accuracy of the decoder,
we perform a point-by-point comparison between the phase shifts extracted using the quadrature Ramsey decoder (Eq.~\ref{decoder}) and the known randomly sampled phase shifts applied to the final $\pi/2$-pulse, as shown in the red points in Fig.~\ref{Fig7}.
The blue points in Fig.~\ref{Fig7} correspond to Monte-Carlo simulations using QPN and the LO noise spectrum as input parameters,
and are in good agreement with the experiment.


\section{Self-comparison clock interrogation with quadrature Ramsey protocol}
\label{Appendix:Cycle}

Before clock interrogation, the atomic population is transferred from $\ket{{}^1S_0, m_F=9/2}\rightarrow\ket{{}^3P_0, m_F=7/2}$ (note as $\ket{g, 9/2}\rightarrow\ket{e, 7/2}$) via a $\pi$-pulse,
followed by a clean-up pulse on resonant with the 461 nm ${}^1S_0\rightarrow{}^1P_1$ transition to blow out residual populations in the ground state.
Quadrature Ramsey spectroscopy is then performed by interrogating the $\ket{e, 7/2}\rightarrow\ket{g,5/2}$ clock transition,
which has a first-order Zeeman shift sensitivity of 564 Hz/G.
To generate an error signal for frequency feedback,
the phase of the final $\pi/2$-pulse is detuned by $\pi/2$ with respect to the phase of the first $\pi/2$-pulse.
After spectroscopy, 
the lattice is adiabatically ramped back up to full depth for readout.
The populations in the ground and excited clock states of both ensembles are read-out in parallel with imaging pulses along the lattice axis,
with scattered photons collected on an EMCCD camera (Andor, iXon-888).
The excitation fraction is extracted through
$P = (N_e-N_{bg})/(N_g+N_e-2N_{bg})$,
where $N_g, N_e$ and $N_{bg}$ are the ground state population,
excited clock state population, and background counts without atoms, respectively.
The excitation fractions from both ensembles are used as inputs to the phase estimator to extract the clock laser detuning,
which is used to feed back to the acousto-optical modulator on the clock laser path that addresses the atomic resonance.

\begin{figure}
\includegraphics[width=0.45\textwidth]{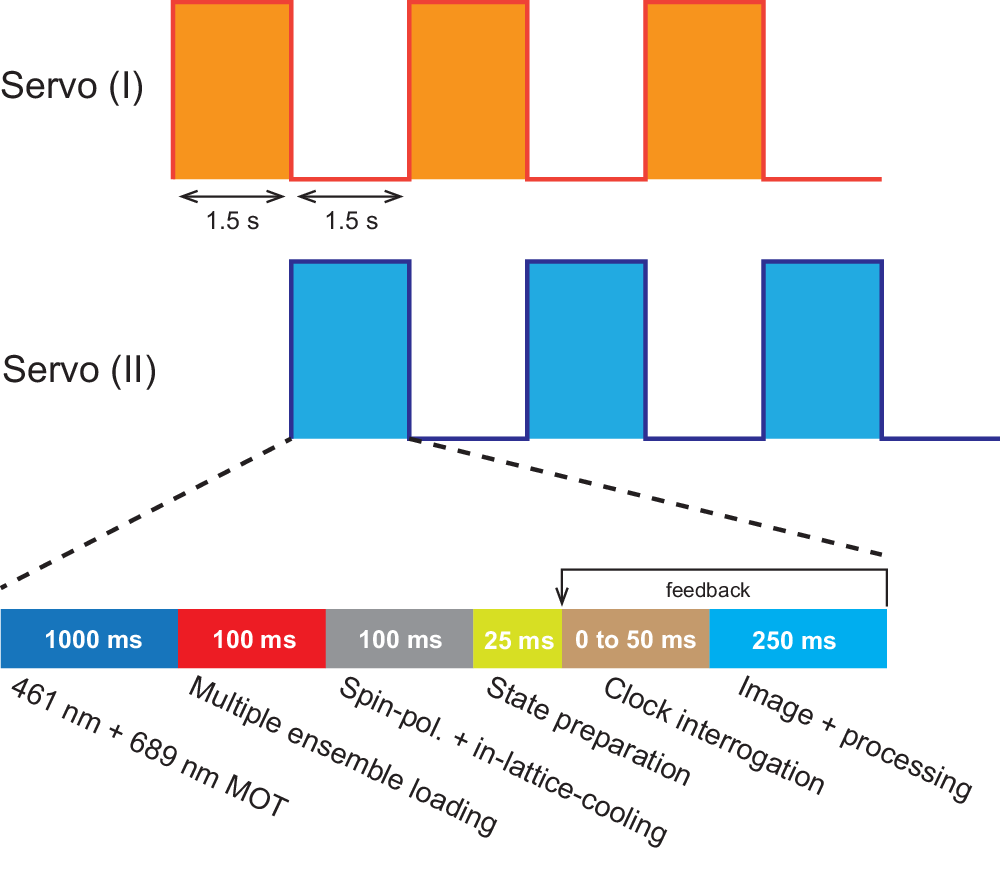}
\caption{\label{Fig8} Experimental sequence for the self-clock comparisons. Two independent atomic servos (I) and (II) are interleaved and compared to determine the clock comparison instability.
The typical experimental cycle is shown in the bottom (not to scale), which takes about 1.5 s, and thus leads to roughly 3 s dead times in each servo.
An additional 250 ms is spent on nuclear spin state initialization when probing four atom ensembles with phase estimation protocol (not shown in the figure).
}
\end{figure}

Due to the lack of an additional independent optical lattice clock to compare against,
in order characterize the clock instability we perform a self-comparison.
Two asynchronous independent servos sharing the same LO are interleaved on a single experiment apparatus. Separate digital servos correct the clock laser frequency independently for each servo and generate a difference frequency signal which is used to extract the Allan deviation and infer the clock instability.
A typical experimental timing sequence for the self-comparison is shown in Fig.~\ref{Fig8}.
The typical dead time is about 1.5 s within each servo,
and thus 3 s per clock cycle when accounting for the other servo.
The primary contribution comes from sample loading,
which takes about 1 s to cool, trap, and optically pump the atoms, with an additional 100 ms required to prepare multiple ensembles rather than a single ensemble.
The camera imaging and data processing takes about 250 ms,
while an additional 250 ms is spent on  the nuclear spin state preparation when performing phase estimation with four atom ensembles, 
which is not shown in Fig.~\ref{Fig8}


\section{Phase estimation and quadrature Ramsey with four atom ensembles}

\subsection{Hyperfine state initialization}

The experimental sequence used to initialize four atom ensembles is shown in Fig.~\ref{Fig3}(a) in the main text.
We elaborate on some of the experimental details here.
First, we start with a sample of spin-polarized atoms in the $\ket{g, 9/2}$ state with four ensembles spaced evenly over a 1 cm extent.
A first $\pi$-pulse on the $9/2-7/2$ transition transfers the population into $\ket{e, 7/2}$,
followed by a clean-up pulse on the 461 nm ${}^1S_0\rightarrow{}^1P_1$ transition.
We then prepare the atoms in a 50:50 superposition of the $\ket{e, 7/2}$ and $\ket{g, 5/2}$ state with a clock $\pi/2$-pulse,
and coherently transfer the population in $\ket{g, 5/2}$ into $\ket{e, 3/2}$ via a $\pi$-pulse on resonance with the $5/2-3/2$ transition.
This creates a superposition of two nuclear spin states in the excited clock state manifold, $\ket{e, 7/2}$ and $\ket{e, 3/2}$,
which freely evolve for roughly $T_p=40$ ms under a linear magnetic field gradient tuned to about 45 mG/cm to maximize the differential phase shifts between the adjacent atom ensembles.
We then coherently transfer the populations and close the interferometer via a $\pi$-pulse on the $7/2-5/2$ transition and a final $\pi/2$-pulse on $5/2-3/2$ transition,
respectively.
Pulses on resonance with either the ${}^1S_0\rightarrow{}^1P_1$ transition or the ${}^3P_{0,2}\rightarrow{}^3S_1$ transitions are combined with clock $\pi$-pulses to clean up residual populations leftover in unwanted states in each pair of ensembles.
$T_p$ and the laser phase of the last $\pi/2$-pulse are fine-tuned such that the four atom ensembles accumulate $0$, $\pi$, $2\pi$, and $3\pi$ phase shifts, respectively,
enabling preparation of two ensemble pairs in distinct nuclear spin states,
pair (1, 3) in $\ket{g, 5/2}$ and pair (2, 4) in $\ket{e, 3/2}$,
using only global pulses without the need for individual ensemble addressing.

Relevant transitions, their differential magnetic moments, and the measured Rabi frequencies used for state initialization and clock interrogation are summarized in Table~\ref{Table1}.

\begin{table}
\caption{\label{Table1}
Relevant transitions used in state initialization of four atom ensembles.}
\begin{ruledtabular}
\begin{tabular}{ccc}
 Transition & B field sensitivity & Rabi frequency$/(2\pi)$ \\
\hline
$\ket{g,9/2}\rightarrow\ket{e,7/2}$ & 194 Hz/G & 110 Hz \\
\hline
$\ket{e,7/2}\rightarrow\ket{g,5/2}$ & 564 Hz/G & 125 Hz \\
\hline
$\ket{g,5/2}\rightarrow\ket{e,3/2}$ & -23 Hz/G & 142 Hz \\
\hline
$\ket{e,3/2}\rightarrow\ket{g,1/2}$ & 340 Hz/G & 166 Hz \\
\end{tabular}
\end{ruledtabular}
\end{table}

\subsection{Phase estimation algorithm}

\label{Appendix:PhaseEstimation}

As described in Sec.~\ref{sec:PE} of the main text, by probing asynchronously on two hyperfine clock transitions,
we obtain two phases from the two quadrature Ramsey measurements: $\theta_A = \mathcal{R}(P_1, P_3)$ and $\theta_B = \mathcal{R}(P_2, P_4) $.
We constrain $\theta_A$ within $[-\pi, \pi]$ by choosing $T_A \leq 50$ ms,
and we have $\theta_A = 2 \pi \Delta f T_A$,
in which $\Delta f$ is the LO detuning.
Given that $T_B = 1.7\times T_A$,
the actual phase shift $\theta_{{\rm act}, B}$ accumulated during $T_B$ may exceed the range of $[-\pi, \pi]$,
which cannot be handled by the quadrature Ramsey decoder (Eq.~\ref{decoder}) due to the $2\pi$ ambiguity.
Consequently, this results in $\theta_B = \theta_{{\rm act}, B} - k \times 2\pi (k = 0, \pm1)$ that could lead to phase slips.

To ascertain $\theta_{{\rm act}, B}$, 
we utilize the following phase estimator to determine the value of $k$ that resolves the $2\pi$ ambiguity.
In the limit that the LO noise is dominated by low frequency noise and for a low duty cycle,
we can assume that the frequency of the LO did not significantly change during the second half of the interrogation period $T_B$, and we approximate the estimated phase shift accumulated by ensembles (2, 4) as $\theta_{\rm est,B}= 1.7 \times \theta_A$.
In our estimator, 
we compare $\theta_{\rm est,B}$ with $\theta_B + k\times 2\pi (k = 0, \pm1)$ and identify the particular value of $k$ that yields the smallest difference.
We then use $\theta_{\rm{act},B} = \theta_B + k \times 2\pi$ to feedback on the LO according to $\Delta f = \theta_{{\rm act},B} /(2 \pi T_B)$.

\section{Scaling of the Dick-noise-limited clock instability}
\label{Appendix:DickScaling}

The instability of a clock limited by Dick noise is given by~\cite{Dick1987}
\begin{equation}
    \sigma^2_{\rm Dick}(\tau) = \frac{1}{\tau}\sum_{m=1}^{\infty}\bigg( \frac{g_m^{c2}}{g_0^2} + \frac{g_m^{s2}}{g_0^2} \bigg)S_{\rm LO}^f(m/T_c),
\end{equation}
where $T_c$ is the cycle time,
$m$ is the $m$-th harmonic,
$S_{\rm LO}^f(m/T_c)$ is the one-sided power spectral density of the relative frequency fluctuations of the LO at Fourier frequencies $m/T_c$, and the parameters $g_0$, $g_m^s$, and $g_m^c$ are defined as:
\begin{equation}
    \begin{split}
        g_m^s & = \frac{1}{T_c}\int^{T_c}_0 g(\xi) {\rm sin} (2\pi m \xi /T_c) d\xi ,  \\
        g_m^c & = \frac{1}{T_c}\int^{T_c}_0 g(\xi) {\rm cos} (2\pi m \xi /T_c) d\xi ,  \\
        g_0  &= \frac{1}{T_c}\int^{T_c}_0 g(\xi) d\xi ,  \\
    \end{split}
\end{equation}
with $g(\xi)$ being the sensitivity function of Ramsey spectroscopy sequence.

In the limit that $T_{\pi/2} \ll T$ (where $T_{\pi/2}$ is the duration of $\pi/2$-pulse),
we can approximate the sensitivity function of a Ramsey spectroscopy with free evolution time $T$ as
\begin{equation*}
    g(\xi) = \begin{cases}
    1& \quad 0 < \xi < T,\\
    0& \quad T < \xi < T_c.\\
\end{cases}
\end{equation*}

The LO noise is modeled as
\begin{equation}
    S_{\rm LO}^f(f) = \sum_{k=-2}^0 b_kf^k,
\label{LO_noise_overall}
\end{equation}
in which $b_0$ corresponds to white frequency noise,
$b_{-1}$ corresponds to flicker frequency noise,
and $b_{-2}$ corresponds to random walk frequency noise.

Here we derive the scaling of Dick-limited clock instability as function of interrogation time $T$ for an arbitrary duty-cycle.
As pointed out in Refs.~\cite{Dick1987,Santarelli1998},
for an LO dominated by its flicker frequency noise floor,
$S^f_{\rm LO}(f)$ = $\sigma_{\rm LO}^2/(2 {\rm ln}2) f^{-1}$,
where $\sigma_{\rm LO}$ is the LO flicker frequency instability,
the Dick noise limited clock instability can be approximated as:
\begin{equation}
    \sigma_{\rm Dick}(\tau)\approx\frac{\sigma_{\rm LO}}{\sqrt{2 {\rm ln} 2}} \sum_{m=1}^{\infty} 
    \sqrt{\frac{g^{s2}_m + g_m^{c2}}{g_0^2}}\sqrt{\frac{T_c}{m\tau}},
    \label{Flicker}
\end{equation}
where $T_c$ is the cycle time.

We then have the parameters $g_{0, 1}$ as
\begin{equation}
\begin{split}
    g_0 & = \frac{1}{T_c}\int^{T_c}_0 g(\xi) d\xi \\
    & = \frac{1}{T_c}\int^{T}_0 d\xi \\
    & = T/T_c,
\end{split}
\end{equation}

\begin{equation}
\begin{split}
    g^s_m & = \frac{1}{T_c}\int^{T_c}_0g(\xi){\rm sin}(2\pi m\xi/T_c)d\xi\\
    & = \frac{1}{T_c}\int^{T}_0 {\rm sin}(2\pi m\xi/T_c)d\xi\\ 
    & = -\frac{1}{2\pi m} \bigg({\rm cos}(2\pi m T/T_c) -1\bigg),
\end{split}
\end{equation}

\begin{equation}
\begin{split}
    g^c_m & = \frac{1}{T_c}\int^{T_c}_0g(\xi){\rm cos}(2\pi m\xi/T_c)d\xi\\
    & = \frac{1}{T_c}\int^{T}_0 {\rm cos}(2\pi m\xi/T_c)d\xi\\ 
    & = \frac{1}{2\pi m} {\rm sin}(2\pi m T/T_c),
\end{split}
\end{equation}
and thus after some algebraic manipulation we have:
\begin{equation}
    \sqrt{\frac{g_m^{s2} + g_m^{c2}}{g_0^2}} = \bigg|\frac{{\rm sin}(\pi m T/T_c)}{\pi m T/T_c}\bigg|.
\end{equation}
Plugging this back into Eq.~\ref{Flicker},
we arrive at
\begin{equation}
    \sigma_{\rm Dick}(\tau) =  \frac{\sigma_{\rm LO}}{\sqrt{2{\rm ln}2}} \sum_{m=1}^{\infty}\bigg|\frac{{\rm sin}(\pi mT/T_c)}{\pi mT/T_c}\bigg| \sqrt{\frac{T_c}{m\tau}}.
    \label{scaling_m}
\end{equation}
It's worth noting that taking only the lowest order term $m=1$,
we have Eq.~\ref{scaling_m} as
\begin{equation}
    \sigma_{{\rm Dick}, m=1}(\tau) = \frac{\sigma_{\rm LO}}{\sqrt{2{\rm ln}2}}\bigg|\frac{{\rm sin}(\pi T/T_c)}{\pi T/T_c}\bigg| \sqrt{\frac{T_c}{\tau}},
    \label{scaling_1}
\end{equation}
which matches the approximation given in Eq.~12 of Ref.~\cite{Santarelli1998}.

We can thus fit our data (as taken in Fig.~\ref{Fig2}(b) and Fig.~\ref{Fig4}(c) in the main text) to the scaling in Eq.~\ref{scaling_m}.
However,
we need to bound the maximum $m_{\rm max}$ in the actual numerical fitting given limited computational resources, and due to the finite bandwidth of the Ramsey sensitivity function.
The cutoff $m_{\rm max}$ can be roughly bounded by the multiples of sine waves within the minimum Ramsey free evolution time $T_{\rm min}$ as
\begin{equation}
    m_{\rm max} = {\rm mod}(n T_c/T_{\rm min}),
\end{equation}
in which we conservatively choose $n = 2$.
In our actual data taking,
we have a typical $\pi/2$-pulse duration of $T_{\pi/2}=1.5$ ms, 
and a minimum $T_{\rm min}=2$ ms.
To account for the finite pulse duration,
we can take the average of sensitivity function $g'(\xi)$ over the entire duration of $(0, T + 2 T_{\pi/2})$,
where we have
\begin{equation*}
\begin{split}
    & g'(\xi) = \\
    &\begin{cases}
    {\rm sin}(\Omega\xi) & \quad 0 < \xi < T_{\pi/2},\\
    1& \quad T_{\pi/2} < \xi < T + T_{\pi/2},\\
    {\rm sin}(\Omega(T + 2T_{\pi/2}-\xi)) & \quad T + T_{\pi/2} < \xi < T + 2T_{\pi/2},\\
\end{cases}
\end{split}
\end{equation*}
where $\Omega T_{\pi/2}= \pi/2$,
with $\Omega$ being the Rabi frequency.
For $T_{\pi/2}=1.5$ ms and $T=2$ ms,
this yields an effective duration of $T'_{\rm min}=3.9$ ms, giving $m_{\rm max} = 1700$ in our numerical fitting.

In order to fit to the instability data shown in both Fig.~\ref{Fig2}(b) and Fig.~\ref{Fig4}(c), we use a simplified expression for the Dick noise that assumes only flicker frequency noise, and treat the instability of the LO as a free parameter. The extracted LO flicker frequency instability of $\sigma_{\textrm{FF}} = 3.2(2)\times10^{-15}$, 
which is roughly a factor of $2-3\times$ larger than the value provided by the manufacturer of the LO (Menlo Systems).
This discrepancy likely arises in part from not including the contributions of random walk and white frequency noise in the fit. 
In addition, the LO is now in a relatively noisy vibration and thermal environment, and we believe its noise spectrum is now degraded (see next section for details).

\section{Reducing clock instability by increasing interrogation times with multiple ensembles}

\subsection{Theoretically expected clock instability}
\label{Appendix:increasing_times}

The expected clock instability is given by the quadrature sum of QPN and Dick noise,
\begin{equation}
    \sigma^2 = \sigma_{\rm QPN}^2 + \sigma^2_{\rm Dick}.
\end{equation}

In particular, QPN scales with the Ramsey interrogation time ($T$) and atom number ($N$, per ensemble) as (Eq.~\ref{QPN})
\begin{equation}
    \sigma_{\rm QPN} \propto \sqrt{\frac{T + T_d}{NT^2}}.
\label{e2}
\end{equation}
The Dick noise limited clock instability scales with $T$ roughly as (Eq.~\ref{scaling_m})
\begin{equation}
    \sigma_{\rm Dick} \propto \sigma_{\rm LO} \sum_{m=1}^{\infty}\bigg|\frac{{\rm sin}(\pi mT/T_c)}{\pi mT/T_c}\bigg| \sqrt{\frac{T_c}{m}},
\label{e3}
\end{equation}
in which $m_{\rm max}\approx 1700$ in our case.
Both Eq.~\ref{e2} and Eq.~\ref{e3} suggest that increasing the interrogation time $T$ reduces the clock instability.

In our case,
the clock instability is Dick noise limited due to the LO noise ($\sigma_{\rm LO}\approx 10^{-15}$) and relatively low duty-cycle ($T/T_c<2$\%),
and thus has no atom number dependence given that $N$ is sufficiently large (4000 atoms per ensemble in quadrature Ramsey with two ensembles, and 2000 atoms per ensemble in phase estimation protocol with four ensembles).
We note that, however, even in the case of QPN limited,
we are still taking advantage of the information from all the atoms and should not suffer from reduced atom numbers with the use of multiple ensembles.

For standard Ramsey spectroscopy with a single ensemble,
the interrogation time is $T_1$, limited by requiring that the phase of the LO remain within the range of $[-\pi/2,\pi/2]$.
For quadrature Ramsey protocol with two ensembles,
the interrogation time is increased to $T_2 = 2\times T_1$ by extending the interpretable LO phase shift to $[-\pi,\pi]$.
Experimentally, we achieve a reduction of 1.36(5) in the measured clock instability that is roughly consistent with theoretically expected reduction in the Dick noise limit (Fig.~\ref{Fig2}(b) in the main text).

For the phase estimation protocol with four ensembles we demonstrate in Fig.~\ref{Fig4},
the interrogation time of the second ensemble pair can be further extended by a factor of 1.7 compared to the case of two ensembles as $T_4 = 1.7\times T_2$,
or a factor 3.4 compared to the case of a single ensemble as $T_4 = 3.4\times T_1$. 
And we achieve a reduction of 1.2(1) in the measured clock instability when compared to the quadrature Ramsey protocol,
and a reduction of 2.08(6) when compared to the standard Ramsey approach,
both are again roughly consistent with the theoretical expectations (Fig.~\ref{Fig4}(c) in the main text).

\subsection{Numerical calculations of the clock instability}

The QPN limited instability ($\sigma_{\rm QPN}$) for Ramsey spectroscopy is given in Eq.~\ref{QPN} in the main text.
To calculate $\sigma_{\rm Dick}$ (Eq.~\ref{scaling_m}),
the LO frequency noise spectrum is modeled with Eq.~\ref{LO_noise_overall},
and the coefficients $b_m$ ($m=0,-1, -2$) are extracted from the measured cavity instabilities provided by the manufacturer by beating the LO against another independent cavity-stabilized laser,
and are given as
\begin{equation}
    \begin{split}
        \sigma_{\rm WH}(\tau) &= 5.3^{+2.2}_{-2.0} \times10^{-16}/\sqrt{\tau},\\
        \sigma_{\rm FF}(\tau) &= 1.3^{+0.2}_{-0.1} \times10^{-15},\\
        \sigma_{\rm RW}(\tau) &= 1.0^{+0.1}_{-0.1} \times10^{-15}\sqrt{\tau},\\
    \end{split}
\end{equation}
where $\sigma_{\rm WH}$, $\sigma_{\rm FF}$, and $\sigma_{\rm RW}$ correspond to the estimated fractional white, flicker, and random walk frequency noise instabilities,
respectively.
These instabilities (at $\tau=1$ s) are then converted into $b_m$ as
\begin{equation}
    \begin{split}
        b_0  = 2\sigma_{\rm WH}^2 &= 5.7^{+3.0}_{-2.5} \times10^{-31} ~{\rm Hz}^{-1},\\
        b_{-1} = \frac{\sigma^2_{\rm FF}}{2{\rm ln}2} &= 1.3^{+0.3}_{-0.3} \times10^{-30},\\
        b_{-2}  =\frac{6\sigma^2_{\rm RW}}{(2\pi)^2} &= 1.6^{+0.4}_{-0.3} \times10^{-31} ~{\rm Hz}.\\
    \end{split}
\label{LO_noise}
\end{equation}

\begin{figure}
\includegraphics[width=0.45\textwidth]{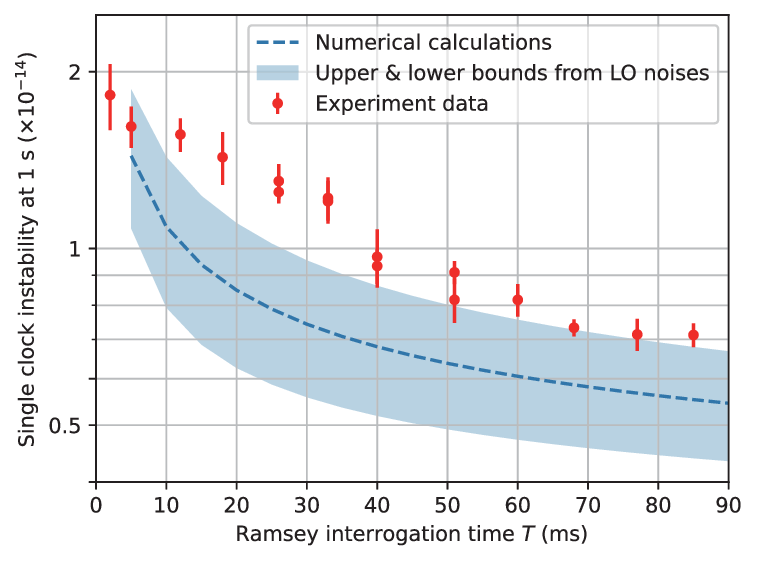}
\caption{\label{Fig9} Numerical calculations of the expected clock instability using the manufacturer's measurements of the LO instability as inputs into Eq.~\ref{LO_noise}.
The shaded region represents upper and lower bound limits from the uncertainty in the LO noise estimates.
The experimental data are taken from Fig.~\ref{Fig4}(c) in the main text,
divided by $\sqrt{2}$ to infer the single clock instability.
The discrepancy between experimental data and numerical calculations is likely because the noise spectrum of our LO is degraded in a relatively noisy vibration and thermal environment.
Error bars correspond to 1$\sigma$ standard deviation.
}
\end{figure}

The calculated single clock instability is shown in Fig.~\ref{Fig9},
where the shaded regions represent uncertainties in the calculation,
primarily arising from the estimated frequency noise spectrum of the LO.
The scatter points represent single clock instability corresponding to the data in Fig.~\ref{Fig4}(c) in the main text,
divided by $\sqrt{2}$ assuming equal contribution from either atomic servo.
The measured instabilities are slightly greater than the numerical calculations using LO frequency noises provided by the manufacturer,
which likely because our LO is now in a relatively noisy vibration and thermal environment, and we believe its noise spectrum is now degraded.

\section{Calculations of phase slip and phase estimation failure probabilities}
\label{Appendix:phase_slip_probability}

\subsection{Generation of LO frequency time trace}

In order to predict the likelihood of a phase slip and the efficacy of the phase estimation algorithms we employ, we first randomly generate a simulated time series of the LO frequency using its power spectral density (PSD) following Ref.~\cite{timetrace_pra_2018},
where the LO PSD is sampled with a discretized frequency $\Delta f$.
The amplitude $A_{\nu}(f)$ is given by
\begin{equation}
    A_{\nu}(f) = e^{i \eta} \sqrt{2 S_{\nu}(f) \Delta f},
\end{equation}
where $S_{\nu}(f)$ is the PSD of frequency noise in units of $\rm{Hz^2 \cdot Hz^{-1}}$ and $\eta$ is a phase randomly drawn from a uniform distribution within $[0, 2\pi]$ at each $f$.
The amplitude is then converted into a time trace $\nu(t)$ via fast Fourier transform
\begin{equation}
    \nu(t) = \mathcal{F}\{ A_{\nu}(f) \}.
\end{equation}

\subsection{Phase slip probability}

\begin{figure}
\includegraphics[width=0.45\textwidth]{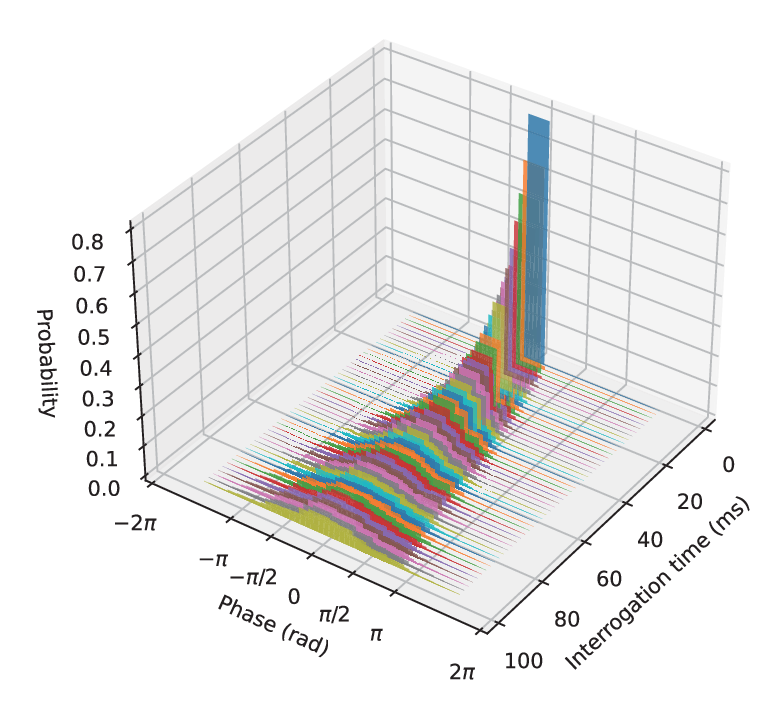}
\caption{\label{Fig10} Distribution of LO phase evolution under various interrogation time $T$. The distribution is computed with a time trace of LO frequency at each $T$ over the course of 1000 measurement cycles.
The time trace is generated with a sampling rate of 10 kHz using our experimental parameters, including the LO noise PSD and a dead time of 3.5 s.
}
\end{figure}

To calculate the phase slip probability,
we first generate a time trace for up to 1000 cycles with a dead time of 3.5 s and a sampling rate of 10 kHz. 
We then compute the phase evolution within $T$ through the integral of the instantaneous frequency for each cycle.
The distributions of phase evolution over 1000 cycles for various $T$ are shown in Fig.~\ref{Fig10}.
Each distribution is then fitted to a Gaussian function to extract the standard deviation $\sigma$.

\begin{figure}
\includegraphics[width=0.45\textwidth]{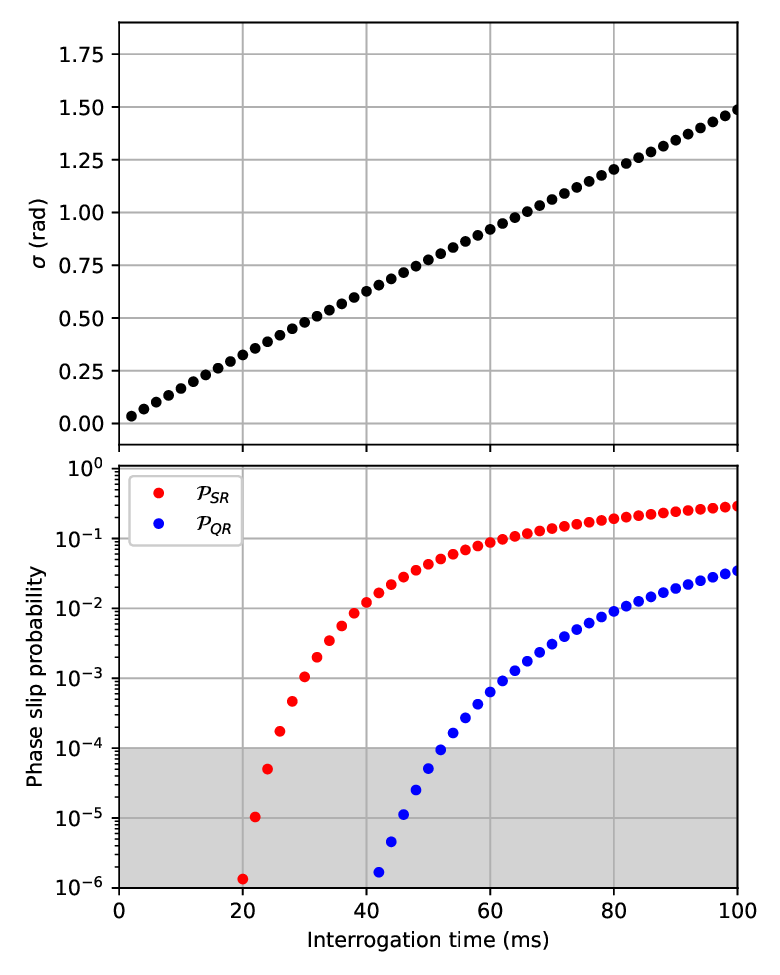}
\caption{\label{Fig11} Calculation of phase slip probability.
Top: Standard deviation $\sigma$ at various interrogation times $T$ extracted via Gaussian fit to the distributions in Fig.~\ref{Fig10}.
Bottom: Computed phase slip probabilities $\mathcal{P_{\rm SR}}$ ($\mathcal{P_{\rm QR}}$) using Eq.~\ref{error_func} under various interrogation times for standard Ramsey (SR, $\Phi_t = \pi/2$) and quadrature Ramsey (QR, $\Phi_t = \pi$), respectively. The shaded area bounds the phase slip probability below $10^{-4}$.
}
\end{figure}

The phase slip probability $\mathcal{P}$ can be computed using the Gaussian error function as
\begin{equation}
    \mathcal{P} = 1- {\rm erf}\bigg(\frac{\Phi_{\rm t}}{\sqrt{2}\sigma}\bigg),
\label{error_func}
\end{equation}
where $\Phi_{\rm t}$ corresponds to $\pi/2$ ($\pi$) for standard (quadrature) Ramsey protocol.
The computed standard deviations and the associated phase slip probabilities for both the SR and QR protocols are shown in Fig.~\ref{Fig11}.
When bounding the phase slip probability below $10^{-4}$,
we find the maximal interrogation times of $26(3)$ and $52(3)$ ms for standard Ramsey and quadrature Ramsey, respectively, consistent with our experimental observations.

We note that measurements of clock instability using two interleaved servos in a single clock are blind to both servos simultaneously fringe hopping,
e.g., the clock laser occasionally takes sudden frequency steps (as opposed to slower drifts of the LO) that would not be detectable in a self-comparison. 
In our apparatus, such a large frequency step would primarily come from excessive external vibrations which can rail the active vibration isolation stage holding the clock reference cavity, causing the cavity to experience a large acceleration.
We monitor several experiment parameters throughout our data taking to rule this out.
First, the AVI stage is monitored throughout the experiment to ensure it stays within its dynamic range.
Second, since we are preparing atoms onto designated states via multiple $\pi/2$- and $\pi$-pulses, any sudden frequency jumps on the order of $>10$ Hz (as compared to the Rabi frequency of $2\pi\times 100$ Hz) would cause substantial reduction in state transfer efficiency for both servos simultaneously.
Third, for quadrature Ramsey, the frequency correction in each servo cycle is bounded within $(-1/2T, +1/2T)$ by the arcsine function. It would require frequency jumps in multiple consecutive experimental cycles to cause a fringe slip (corresponding to multiples of $|1/T|$). The recorded corrections in each cycle are also monitored throughout the experiment.

\subsection{Failure probability of the quadrature Ramsey with phase estimation protocol}

\begin{figure}
\includegraphics[width=0.45\textwidth]{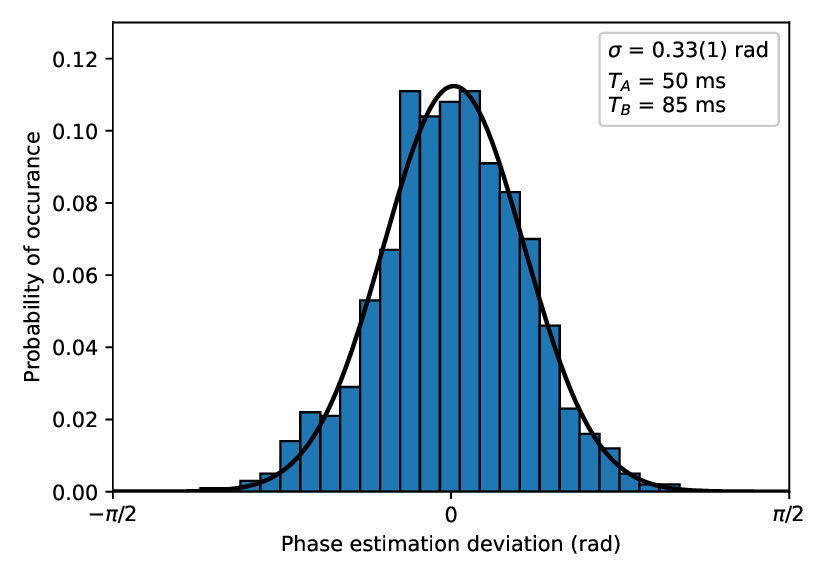}
\caption{\label{Fig12} Monte-Carlo simulation of phase estimation failure probability.
Plot shows the simulated distribution of phase estimation deviations computed over 1000 cycles under $T_A$ = 50 ms and $T_B$ = 85 ms (the same interrogation times as experimentally demonstrated in Fig.~\ref{Fig4}(c)).
The black curve is a Gaussian fit to the distribution, resulting in a standard deviation of $\sigma =0.33(1)$ rad,
corresponding to a probability of $\mathcal{P_{\rm PE}}<10^{-10}$ using Eq.~\ref{error_func} ($\Phi_t = \pi$).
This, when accounting for the phase slip probability during $T_A=50$ ms,
results in a failure probability of $\mathcal{P_{\rm QR \& PE}} < 10^{-4}$,
consistent with our experiment observation.}
\end{figure}

The failure probability for our four-ensemble quadrature Ramsey with phase estimation protocol is given by
\begin{equation}
    \mathcal{P_{\rm QR \& PE}} = 1-(1-\mathcal{P_{\rm QR,A}})(1-\mathcal{P_{\rm PE}}),
\end{equation}
where $P_{\rm QR,A}$ is the quadrature Ramsey phase slip probability during the shorter interrogation period $T_A$,
and $\mathcal{P_{\rm PE}}$ is the phase estimation failure probability.

To compute $\mathcal{P_{\rm PE}}$,
we perform Monte-Carlo simulations by generating a LO frequency time trace over 1000 cycles.
In each cycle, we extract the actual accumulated phase shift $\theta_{{\rm act}, A}$ during $T_A$,
which yields a guessed phase shift at $T_B$ as $\theta_{{\rm est}, B} = \theta_{{\rm act}, A}\times (T_B/T_A)$,
assuming a constant LO detuning.
$\theta_{{\rm est}, B}$ is then compared with the actual phase shift $\theta_{{\rm est}, B}$ during $T_B$ to determine the phase estimation deviation as
\begin{equation}
\Delta\theta_{\rm PE} = \theta_{\rm B} - \theta_{\rm est, B},
\end{equation}
where the phase estimation breaks down when $|\Delta\theta_{\rm PE}| > \pi$.

Fig.~\ref{Fig12} shows a distribution of the phase estimation deviations computed over 1000 cycles under $T_A = 50$ ms and $T_B = 85)$ ms,
which is experimentally demonstrated in Fig.~\ref{Fig4}(c).
A Gaussian fit yields in a standard deviation of 0.33(1) rad,
corresponding to $\mathcal{P_{\rm PE}} < 10^{-10}$ calculated using Eq.~\ref{error_func} (with $\Phi_t =\pi$).
After accounting for the phase slip probability during $T_A = 50$ ms ($\mathcal{P_{\rm QR,A}} \simeq 0.8\times10^{-4}$),
we find a failure probability of $\mathcal{P_{\rm QR \& PE}} < 1\times10^{-4}$.
We note that due to the requirement of $(2k+1)\pi/2$ phase shifts $(k=0,1,2\dots)$,
the next available $T_B$ at $T_A = 50$ ms is 255 ms,
which has a phase estimation failure probability of $>10\%$ and is thus not experimentally feasible.

\section{Monte Carlo simulation for state-of-the-art LOs}
\label{Appendix:HigherQualityLO}

\begin{figure}
\includegraphics[width=0.45\textwidth]{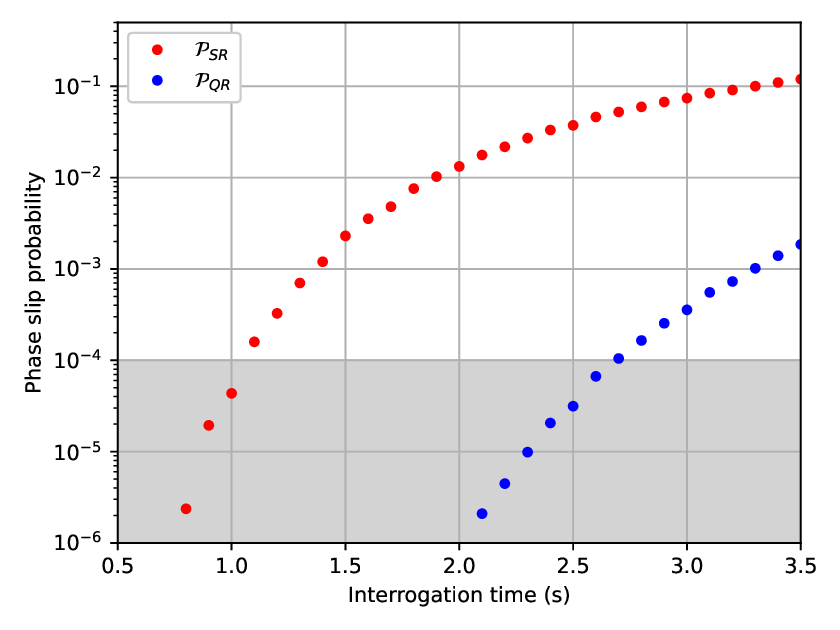}
\caption{\label{Fig13} Monte Carlo simulations of phase slip probabilities for state-of-the-art LOs.
Computed phase slip probabilities $\mathcal{P_{\rm SR}}$ ($\mathcal{P_{\rm QR}}$ ) under various interrogation times for standard Ramsey (SR, $\Phi_t = \pi/2$) and quadrature Ramsey (QR, $\Phi_t = \pi$), respectively.
A fixed dead time of 600 ms and the flicker noise floor of state-of-the-art LOs ($4\times10^{-17}$) are used as input parameters.
The shaded area bounds the phase slip probability below $10^{-4}$.
A factor of 2 enhancement in coherence times is feasible with the QR protocol.
}
\end{figure}

To connect our multiple-ensemble protocols to the state-of-the-art optical lattice clocks that make use of higher quality LOs~\cite{kessler_SiCavity_2012,Matei_SiCavity_2017,Zhang_SiCavity_2017,Robinson_SiCavity_2019,Kedar_SiCavity_2023},
we perform Monte Carlo simulations to predict the phase slip probabilities for these systems.
In our simulations, we assume LOs with a flicker noise floor of $<4\times10^{-17}$ (coherence time $>600$ ms) and a duty-cycle above 50\%~\cite{mcgrew_atomic_2018,Oelker_ClockStability_2019}.
Similar to Appendix~\ref{Appendix:phase_slip_probability},
we calculate the phase slip probability as a function of interrogation time under a fixed dead time of 600 ms and a flicker noise floor of $4\times10^{-17}$.
Fig.~\ref{Fig13} shows the simulated phase slip probabilities for standard Ramsey (SR) and quadrature Ramsey (QR) approaches, respectively, for state-of-the-art LOs.
Bounding the phase slip probabilities below $10^{-4}$,
we predict that a factor of 2 enhancement in coherence times with QR protocol is feasible for state-of-the-art LOs with a duty-cycle $>50\%$.

\begin{figure}
\includegraphics[width=0.45\textwidth]{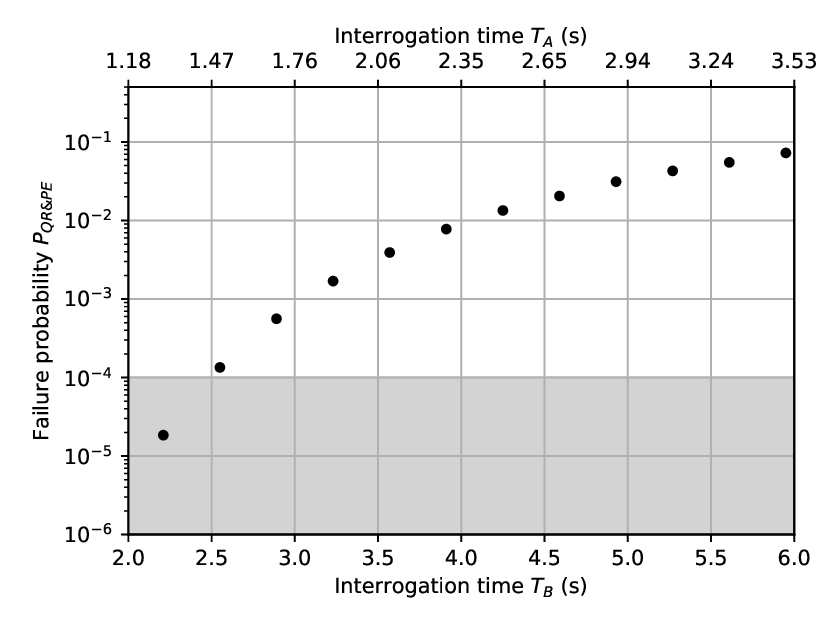}
\caption{\label{Fig14} Simulated failure probability for state-of-the-art LOs.
We assume a flicker noise floor of $4\times10^{-17}$, a fixed dead time of 600 ms for interrogation period $T_B$,
and we choose $T_B = 1.7\times T_A$.
The black scatter points correspond to the failure probability $\mathcal{P_{\rm QR\& PE}}$.
By bounding $\mathcal{P_{\rm QR\&PE}}$ below $10^{-4}$,
we find a maximal interrogation time $T_B$ of around 2.45 s with four-ensemble QR \& PE protocol.
}
\end{figure}

We then compute the failure probability four-ensemble quadrature Ramsey and phase estimation (QR \& PE) protocol with higher quality LOs,
as shown in the black scatter points Fig.~\ref{Fig14}.
We assume a flicker noise floor of $4\times10^{-17}$, a fixed dead time of 600 ms for interrogation period $T_B$,
and we choose $T_B = 1.7\times T_A$.
By bounding $\mathcal{P_{\rm QR\&PE}}$ below $10^{-4}$,
we find a maximal interrogation time $T_B$ of around 2.45 s with four-ensemble QR \& PE protocol.
As the four-ensemble QR \& PE protocol doesn't improve over the two-ensemble QR protocol (interrogation time of around 2.7 s) for higher-quality LOs and with $>50$\% duty-cycles, the simulations confirm the intuition and arguments presented in Section~\ref{sec:Discussion} of the main text
that a third pair of ensembles would be needed to keep track of the phase evolution during $T_B - T_A$ for a higher duty cycle clock and benefit from multi-ensemble phase estimation with $>2$ ensembles.

\section{Potential systematics associated with multiple-ensemble protocols}
\label{Appendix:Systematics}

In this section,
we consider potential sources of additional systematic shifts arising from the quadrature Ramsey spectroscopy and phase estimation approaches using multiple ensembles,
and strategies to mitigate such effects.

\subsection{Magnetic field gradient}

In this work we employ a magnetic field gradient to introduce a $\pi/2$ differential phase shift between the ensemble pairs,
which is on the order of 20 mG/cm depending on the interrogation time $T$ and magnetic moment of clock transition being used.
Under a bias magnetic field of $\sim5.5$ G,
the magnitude of first-order Zeeman shift is about 3.1 kHz for the $\ket{e, 7/2}\rightarrow\ket{g, 5/2}$ transition,
and is about 1.87 kHz for the $\ket{e, 3/2}\rightarrow\ket{g, 1/2}$ transition.
While the first-order Zeeman shift is typically cancelled out by averaging transitions between opposite spin states,
to support sufficient suppression the magnetic field needs to be stable within $\pm$0.1 mG between experimental cycles for a coherence time of 100 ms,
estimated using 10\% of the Fourier-limited linewidth.

The addition of a magnetic field gradient also contributes to the second-order Zeeman shift as well as potential differential tensor Stark shifts due to spatially varying B-field vectors across the lattice axis.
As pointed out in Appendix~\ref{Appendix:DifferentialPhaseShift}, 
we found a fractional frequency gradient of $-1.2(2)\times10^{-18}$/cm at our operational depth of 15~$E_{\rm rec}$.
In future work,
we plan to mitigate this effect by employing a more uniform linear gradient through the addition of Golay coils~\cite{Golay_coils_1958},
requiring an upgrade to our experimental apparatus that is currently underway.

For optical lattice clocks with higher quality LOs that enable coherence times approaching second-scale,
a smaller magnetic field gradient and a magnetically less sensitivity clock transition can be used to generate the $\pi/2$ differential phase shift for quadrature Ramsey spectroscopy.
For example,
with the $5/2-3/2$ transition,
which has a magnetic field sensitivity of $-22.4$ Hz/G,
only a magnetic field gradient of 3~mG/cm is needed at $T=3$ s interrogation time for a pair of ensembles separated by 1 mm~\cite{bothwell_resolving_2022}.
In addition, spatially-resolved read-out can be carried out with high-resolution imaging to better characterize the gradient, as was demonstrated in Refs.~\cite{bothwell_resolving_2022,campbell_2017,marti_2018}, which can then be used to account for and subtract off the inhomogeneous shifts arising from the gradient.

\subsection{Systematic shifts from off-resonant clock pulses}
\label{Appendix:ProbeShift}
The additional clock $\pi/2$-pulses employed in the four-ensemble phase estimation protocol will lead to additional light shifts on the off-resonant ensembles.
Given the measured sensitivity of -13(2) Hz (W cm${}^{-2}$)${}^{-1}$ from prior work~\cite{probe_light_shift},
we estimate a probe light shift of $6\times10^{-16}$ caused by the on-resonant clock pulse with a Rabi frequency of $2\pi\times140$ Hz.
This sets an upper bound limit below $1\times10^{-15}$ for the shift arising from the off-resonant clock pulses, after accounting for the detuning and Clebsh-Gordan coefficient.
While this is a sizable systematic shift for the state-of-the-art clocks,
given the large Rabi frequency employed in this work,
we note this effect can be reduced by either lengthening the clock pulses,
or through the adoption of more sophisticated composite pulse sequences such as hyper-Ramsey spectroscopy~\cite{hyper_ramsey}. 

\subsection{Servo errors from phase decoder}

The phase decoder we employ here for quadrature Ramsey spectroscopy assumes a differential phase shift of $\Delta\phi=\pi/2$. Inaccuracy in $\Delta\phi$
could introduce a frequency offset in the phase decoder (Eq.~\ref{decoder} in the main text).
In the limit of $\delta\phi\ll\Delta\phi$,
the phase shift error $\delta\phi$ will be converted into a decoder error of $\delta\phi/2$,
and thus result in a frequency offset of $\delta\phi/2\pi T$.

We note that the phase gradient across the lattice is $\pi/2$/cm, which means that the phase difference across an ensemble ($1/e^2$ radius of 500 $\mu$m) is on the order of 100 mrad (between its top and bottom). 
However, as we perform a spatial average over each ensemble, to first order the phase gradient across each individual ensemble will contribute to decoherence within each ensemble rather than inaccuracy in the differential phase shift. 
The dominant source of error will instead be inaccuracy in the differential phase shift between ensemble pairs rather than the phase gradient across each individual ensemble.

In our case, the differential phase error primarily comes from drifts in the magnetic field gradient and any uncertainty in separation between center-of-mass of the ensembles.
We note that the magnetic field gradient was characterized with synchronous differential comparison in prior works~\cite{zheng_differential_2022,zheng_redshift_2023},
demonstrating long-term differential instability and inaccuracy at $10^{-19}$ level.
The ensemble separations are well controlled with the data acquisition card with a timing precision of 100 ns (10 MHz clock reference), corresponding to a separation error less than $<0.1~\mu$m (compared to cm-scale separation).

Assuming a phase error of 10 mrad,
this would lead to an error on the order of $1\times10^{-17}$ at an interrogation time of $T=100$ ms,
which could be further suppressed well below the $10^{-18}$ level using longer interrogation times with a better LO. In addition, averaging interleaved measurements on opposite hyperfine transitions, as is typically employed to cancel the first order Zeeman shift, will to lowest order also cancel this error. Finally, measurements of $\delta\phi$ can be performed with higher precision if necessary by taking advantage of the ability to probe well beyond the LO coherence time using synchronous differential comparisons \cite{zheng_differential_2022,bothwell_resolving_2022,zheng_redshift_2023}.

\subsection{Cross talk and line pulling}

We observe cross talk between the two hyperfine clock transitions used for the four-ensemble phase estimator at the level of roughly 3\% (Fig.~\ref{Fig3}(c)), which could potentially contribute to inaccuracy.
We note that line pulling from off-resonant excitation of unintended Zeeman transitions is not unique to our protocol, and the same considerations apply to existing optical lattice clocks~\cite{Falke_PTB_2014,mcgrew_atomic_2018,bothwell_jila_2019,Oelker_ClockStability_2019,PTB_transportable_2017},
whether or not other hyperfine clock transitions are employed.
We also note that the 3\% cross talk in our current demonstration is a technical issue,
primarily limited by clock pulse infidelity,
rather than a fundamental limit. 
The maximum off-resonant excitation is given by the ratio of Clebsch-Gordon coefficients squared and the off-resonant Rabi frequency as~\cite{bothwell_jila_2019}
\begin{equation}
\gamma_{\rm CG} \times  \frac{\Omega^2}{\Omega^2 + (2\pi \Delta\nu_{\rm split})^2},
\end{equation}
where $\gamma_{\rm CG}$ corresponds to the ratio of Clebsch-Gordon coefficient squared,
$\Omega$ is the Rabi frequency, and $\Delta\nu_{\rm split}$ is the splitting between nearby transitions.
There are multiple ways to mitigate this effect:
a) reducing the Rabi frequency of the clock pulse;
b) employing a larger bias magnetic field;
c) improving the initial state preparation to reduce imperfections in spin-polarization and thus cross talk.
With these improvements, we expect it will be feasible to achieve cross talk below 0.1\%, and to characterize and control the corresponding systematic uncertainty arising from line-pulling at below the $10^{-18}$ level.

\subsection{Residual tensor Stark shifts due to multiple $m_F$ states}

For standard optical lattice clocks operating with transitions with a single $m_F$ state,
the lattice laser frequency is often detuned such that the scalar component and tensor components null out the lattice Stark shift~\cite{brown_prl_2017,katori_lattice_2018,kim_lattice_2023,ptb_latticeLight_2023}.
However, 
for our four-ensemble phase estimation protocol, which makes use of two hyperfine clock transitions with multiple $m_F$ states,
such as the $\ket{e,7/2}\rightarrow \ket{g,5/2}$ and $\ket{e,3/2}\rightarrow \ket{g,1/2}$ transitions as used in this work,
the lattice stark shift cannot be nulled out for both transitions at a single lattice frequency.
Here we estimate the magnitude of this residual shift using the tensor coefficient of -0.0058(23) mHz/$E_{\rm rec}$ from prior works~\cite{westergaard_tensor_2011,boyd_thesis_2007},
which yields a residual tensor Stark shift of $2.6(1.0)\times10^{-17}$ between the two clock transitions at 15 $E_{\rm rec}$,
where the uncertainty is limited by knowledge of the tensor coefficient.
However, we would like to point out that the accuracy of the clock need not be limited by the systematic uncertainty of the transition used to track the phase for shorter interrogation times.
The only requirement is that the tensor Stark shift of the transition ($\ket{e, 7/2}\rightarrow\ket{g, 5/2}$ in our case) that has not been nulled should never be allowed to drift off by enough to result in a phase shift on the order of $\pm\pi$, and the lattice frequency can be set to null the tensor Stark shift for the clock transition used for the longer interrogation time ($\ket{e, 3/2}\rightarrow\ket{g, 1/2}$ in our case) and to actually lock the LO.

\subsection{Running wave contamination due to lattice beam intensity mismatch}

To prepare multiple atom ensembles, 
we employ two double-passing acousto-optical modulators (AOMs) to detune the frequency of the retro-reflected lattice laser beam~\cite{zheng_differential_2022}.
This results in approximately 50\% in an intensity mismatch between the incoming and retro-reflected lattice laser beams due to the diffraction efficiency (90\%) of the acousto-optic modulator (AOM), optical path loss and imperfect retro-reflection.
The mismatch could lead to a running wave that introduces additional light shifts,
which would require precise modeling to constrain at the level of $10^{-18}$~\cite{Runningwave_PRA_2019}.
We note that, however, this systematic effect is not new or unique to our system.
For instance, McGrew et al.~\cite{mcgrew_atomic_2018} constrains the systematic uncertainty associated with a 15\% intensity mismatch to below $1\times10^{-19}$.
And we expect that this mismatch could be further mitigated with the use of two phase-referenced, counter-propagating lattice laser beams~\cite{Ushijima_cryogenic_2015} or a bow-tie cavity~\cite{takamoto_test_2020},
by doing a handoff of the ensembles from a loading and transportation lattice to a balanced spectroscopy lattice, or by loading multiple ensembles by moving the location of the MOT field zero rather than moving the lattice.


\begin{thebibliography}{0}%
\makeatletter
\providecommand \@ifxundefined [1]{%
 \@ifx{#1\undefined}
}%
\providecommand \@ifnum [1]{%
 \ifnum #1\expandafter \@firstoftwo
 \else \expandafter \@secondoftwo
 \fi
}%
\providecommand \@ifx [1]{%
 \ifx #1\expandafter \@firstoftwo
 \else \expandafter \@secondoftwo
 \fi
}%
\providecommand \natexlab [1]{#1}%
\providecommand \enquote  [1]{``#1''}%
\providecommand \bibnamefont  [1]{#1}%
\providecommand \bibfnamefont [1]{#1}%
\providecommand \citenamefont [1]{#1}%
\providecommand \href@noop [0]{\@secondoftwo}%
\providecommand \href [0]{\begingroup \@sanitize@url \@href}%
\providecommand \@href[1]{\@@startlink{#1}\@@href}%
\providecommand \@@href[1]{\endgroup#1\@@endlink}%
\providecommand \@sanitize@url [0]{\catcode `\\12\catcode `\$12\catcode `\&12\catcode `\#12\catcode `\^12\catcode `\_12\catcode `\%12\relax}%
\providecommand \@@startlink[1]{}%
\providecommand \@@endlink[0]{}%
\providecommand \url  [0]{\begingroup\@sanitize@url \@url }%
\providecommand \@url [1]{\endgroup\@href {#1}{\urlprefix }}%
\providecommand \urlprefix  [0]{URL }%
\providecommand \Eprint [0]{\href }%
\providecommand \doibase [0]{https://doi.org/}%
\providecommand \selectlanguage [0]{\@gobble}%
\providecommand \bibinfo  [0]{\@secondoftwo}%
\providecommand \bibfield  [0]{\@secondoftwo}%
\providecommand \translation [1]{[#1]}%
\providecommand \BibitemOpen [0]{}%
\providecommand \bibitemStop [0]{}%
\providecommand \bibitemNoStop [0]{.\EOS\space}%
\providecommand \EOS [0]{\spacefactor3000\relax}%
\providecommand \BibitemShut  [1]{\csname bibitem#1\endcsname}%
\let\auto@bib@innerbib\@empty
\end{thebibliography}%


\begin{thebibliography}{65}

\bibitem{ludlow_optical_2015}
A. D. Ludlow, M. M. Boyd, J. Ye, E. Peik, P. O. Schmidt,
\textit{Optical Atomic Clocks},
Rev. Mod. Phys.
\textbf{87}, 637--701
(2015)

\bibitem{SingleIon_-18_2016}
N. Huntemann, C. Sanner, B. Lipphardt, C. Tamm and E. Peik,
\textit{Single-Ion Atomic Clock with $3\times{10}^{-18}$ Systematic Uncertainty},
Phys. Rev. Lett.
\textbf{116}, 063001
(2016)

\bibitem{mcgrew_atomic_2018}
W. F. McGrew, X. Zhang, R. J. Fasano, S. A. Schäffer, K. Beloy, D. Nicolodi, R. C. Brown, N. Hinkley, G. Milani, M. Schioppo, T. H. Yoon and A. D. Ludlow,
\textit{Atomic Clock Performance Enabling Geodesy Below the Centimetre Level},
Nature
\textbf{564}, 87--90
(2018)

\bibitem{bothwell_jila_2019}
T. Bothwell, D. Kedar, E. Oelker, J. M. Robinson, S. L. Bromley, W. L. Tew, J. Ye and C. J. Kennedy,
\textit{JILA SrI Optical Lattice Clock with Uncertainty of $2.0\times10^{-18}$},
Metrologia
\textbf{56}, 065004
(2019)

\bibitem{Brewer_IonClock_2019}
S. M. Brewer, J.-S. Chen, A. M. Hankin, E. R. Clements, C. W. Chou, D. J. Wineland, D. B. Hume and D. R. Leibrandt,
\textit{$^{27}{\rm Al}^{+}$ Quantum-Logic Clock with a Systematic Uncertainty Below ${10}^{-18}$},
Phys. Rev. Lett.
\textbf{123}, 033021
(2019)

\bibitem{Oelker_ClockStability_2019}
E. Oelker, R. B. Hutson, C. J. Kennedy, L. Sonderhouse, T. Bothwell, A. Goban, D. Kedar, C. Sanner, J. M. Robinson, G. E. Marti, D. G. Matei, T. Legero, M. Giunta, R. Holzwarth, F. Riehle, U. Sterr and J. Ye,
\textit{Demonstration of $4.8\times10^{-17}$ Stability at 1 s for Two Independent Optical Clocks}
Nat. Photon.
\textbf{13}, 714--719
(2019)

\bibitem{chou_optical_2010}
C. W. Chou, D. B. Hume, T. Rosenband and D.J. Wineland,
\textit{Optical Clocks and Relativity},
Science
\textbf{329}, 1630--1633
(2010)

\bibitem{derevianko_hunting_2014}
A. Derevianko and M. Pospelov,
\textit{Hunting for Topological Dark Matter With Atomic Clocks},
Nat. Phys.
\textbf{10}, 933--936
(2014)

\bibitem{takano_geopotential_2016}
T. Takano, M. Takamoto, I. Ushijima, N. Ohmae, T. Akatsuka, A. Yamaguchi, Y. Kuroishi, H. Munekane, B. Miyahara and H. Katori,
\textit{Geopotential Measurements with Synchronously Linked Optical Lattice Clocks},
Nat. Photon.
{\bf 10}, 662--666
(2016)

\bibitem{kolkowitz_gravitational_2016}
S. Kolkowitz, I. Pikovski, N. Langellier, M. D. Lukin, R. L. Walsworth and J. Ye,
\textit{Gravitational Wave Detection with Optical Lattice Atomic Clocks},
Phys. Rev. D
\textbf{94}, 124043
(2016)

\bibitem{PTB_transportable_2017}
S. B. Koller, J. Grotti, St. Vogt, A. Al-Masoudi, S. Dörscher, S. Häfner, U. Sterr and Ch. Lisdat,
\textit{Transportable Optical Lattice Clock with $7\times10^{-17}$ Uncertainty}
Phys. Rev. Lett.
\textbf{118}, 073601
(2017)

\bibitem{grotti_geodesy_2018}
J. Grotti, S. Koller, S. Vogt, S. Häfner, U. Sterr, C. Lisdat, H. Denker, C. Voigt, L. Timmen, A. Rolland, F. N. Baynes, H. S. Margolis, M. Zampaolo, P. Thoumany, M. Pizzocaro, B. Rauf, F. Bregolin, A. Tampellini, P. Barbieri, M. Zucco, G. A. Costanzo, C. Clivati, F. Levi and D. Calonico,
\textit{Geodesy and Metrology with a Transportable Optical Clock},
Nat. Phys.
\textbf{14}, 437--441
(2018)

\bibitem{safronova_search_2018}
M. S. Safronova, D. Budker, D. DeMille, D. F. Jackson Kimball, A. Derevianko and C. W. Clark,
\textit{Search for New Physics with Atoms and Molecules},
Rev. Mod. Phys.
\textbf{90}, 025008
(2018)

\bibitem{takamoto_test_2020}
M. Takamoto, I. Ushijima, N. Ohmae, T. Yahagi, K. Kokado, H. Shinkai and H. Katori,
\textit{Test of General Relativity by a Pair of Transportable Optical Lattice Clocks},
Nat. Photon.
\textbf{14}, 411--415
(2020)

\bibitem{kennedy_dark_2020}
C. J. Kennedy, E. Oelker, J. M. Robinson, T. Bothwell, D. Kedar, W. R. Milner, G. E. Marti, A. Derevianko and J. Ye,
\textit{Precision Metrology Meets Cosmology: Improved Constraints on Ultralight Dark Matter from Atom-Cavity Frequency Comparisons},
Phys. Rev. Lett.
\textbf{125}, 201302
(2020)

\bibitem{bothwell_resolving_2022}
T. Bothwell, C. J. Kennedy, A. Aeppli, D. Kedar, J. M. Robinson, E. Oelker, A. Staron and J. Ye,
\textit{Resolving the Gravitational Redshift Across a Millimetre-Scale Atomic Sample},
Nature
\textbf{602}, 420--424
(2022)

\bibitem{Dick1987}
G. J. Dick,
\textit{LO Induced Instabilities in Trapped Ion Frequency Standards},
Proc. Precise Time and Time Interval Meeting (ed. Sydnor, R. L.)
133--147
(1987)

\bibitem{Santarelli1998}
G. Santarelli, C. Audoin, A. Makdissi, P. Laurent, G. J. Dick and A. Clairon,
\textit{Frequency Stability Degradation of an Oscillator Slaved to a Periodically Interrogated Atomic Resonator},
IEEE Trans. Ultra. Ferro. Freq. Cont.
\textbf{45}, 887--894
(1998)

\bibitem{kessler_SiCavity_2012}
T. Kessler, C. Hagemann, C. Grebing, T. Legero, U. Sterr, F. Riehle, M. J. Martin, L. Chen and J. Ye,
\textit{A Sub-40-mHz-Linewidth Laser Based on a Silicon Single-Crystal Optical Cavity},
Nat. Photon.
\textbf{6}, 687--692
(2012)

\bibitem{Matei_SiCavity_2017}
D. G. Matei, T. Legero, S. Häfner, C. Grebing, R. Weyrich, W. Zhang, L. Sonderhouse, J. M. Robinson, J. Ye, F. Riehle and U. Sterr,
\textit{1.5 $\mu$m Lasers with Sub-10 mHz Linewidth},
Phys. Rev. Lett.
\textbf{118}, 263202
(2017)

\bibitem{Zhang_SiCavity_2017}
W. Zhang, J. M. Robinson, L. Sonderhouse, E. Oelker, C. Benko, J. L. Hall, T. Legero, D. G. Matei, F. Riehle, U. Sterr and J. Ye,
\textit{Ultrastable Silicon Cavity in a Continuously Operating Closed-Cycle Cryostat at 4 K},
Phys. Rev. Lett.
\textbf{119}, 243601
(2017)

\bibitem{Robinson_SiCavity_2019}
J. M. Robinson, E. Oelker, W. R. Milner, W. Zhang, T. Legero, D. G. Matei, F. Riehle, U. Sterr and J. Ye,
\textit{Crystalline Optical Cavity at 4 K with Thermal-Noise-Limited Instability and Ultralow Drift},
Optica
\textbf{6}, 240--243
(2019)

\bibitem{Kedar_SiCavity_2023}
D. Kedar, J. Yu, E. Oelker, A. Staron, W. Milner, J.M. Robinson, T. Legero, F. Riehle, U. Sterr, and J. Ye,
\textit{Frequency stability of cryogenic silicon cavities with semiconductor crystalline coatings},
Optica
\textbf{10}, 464 
(2023)

\bibitem{takamoto_beyond_2011}
M. Takamoto, T. Takano, and H. Katori,
\textit{Frequency Comparison of Optical Lattice Clocks Beyond the Dick Limit},
Nat. Photon.
\textbf{5}, 288--292
(2011)

\bibitem{schioppo_ultrastable_2017}
M. Schioppo, R. C. Brown, W. F. McGrew, N. Hinkley, R. J. Fasano, K. Beloy, T. H. Yoon, G. Milani, D. Nicolodi, J. A. Sherman, N. B. Phillips, C. W. Oates and A. D. Ludlow,
\textit{Ultrastable Optical Clock with Two Cold-Atom Ensembles},
Nat. Photon.
\textbf{11}, 48--52
(2017)

\bibitem{Clements_IonCoherence_2020}
E. R. Clements, M. E. Kim, K. Cui, A. M. Hankin, S. M. Brewer, J. Valencia, J.-S. Chen, C.-W. Chou, D. R. Leibrandt and D. B. Hume,
\textit{Lifetime-Limited Interrogation of Two Independent $^{27}{\mathrm{Al}}^{+}$ Clocks Using Correlation Spectroscopy},
Phys. Rev. Lett.
\textbf{125}, 243602
(2020)

\bibitem{zheng_differential_2022}
X. Zheng, J. Dolde, V. Lochab, B. N. Merriman, H. Li and S. Kolkowitz,
\textit{Differential Clock Comparisons with a Multiplexed Optical Lattice Clock},
Nature
\textbf{602}, 425--430
(2022)

\bibitem{DD_2020}
S. Dörscher, A. Al-Masoudi, M. Bober, R. Schwarz, R. Hobson, U. Sterr and C. Lisdat,
\textit{Dynamical Decoupling of Laser Phase Noise in Compound Atomic Clocks},
Commun. Phys.
\textbf{3}, 185 
(2020)

\bibitem{QND_PRX_2020}
W. Bowden, A. Vianello, I. R. Hill, M. Schioppo and R. Hobson,
\textit{Improving the Q Factor of an Optical Atomic Clock Using Quantum Nondemolition Measurement},
Phys. Rev. X
\textbf{10}, 041052
(2020)

\bibitem{Kim_coherenceAtomicSpecies_2022}
M. E. Kim, W. F. McGrew, N. V. Nardelli, E. R. Clements, Y. S. Hassan, X. Zhang, J. L. Valencia, H. Leopardi, D. B. Hume, T. M. Fortier, A. D. Ludlow and D. R. Leibrandt,
\textit{Improved Interspecies Optical Clock Comparisons through Differential Spectroscopy},
Nat. Phys.
\textbf{19}, 25--29
(2022)

\bibitem{Pedrozo_entanglementclock_2020}
E. Pedrozo-Peñafiel, S. Colombo, C. Shu, A. F. Adiyatullin, Z. Li, E. Mendez, B. Braverman, A. Kawasaki, D. Akamatsu, Y. Xiao and V. Vuletić,
\textit{Entanglement on an Optical Atomic-Clock Transition},
Nature
\textbf{588}, 414--418
(2020)

\bibitem{Robinson_entanglement_2022}
J. M. Robinson, M. Miklos, Y. M. Tso, C. J. Kennedy, T. Bothwell, D. Kedar, J. K. Thompson and J. Ye,
\textit{Direct Comparison of Two Spin Squeezed Optical Clocks Below the Quantum Projection Noise Limit},
arXiv:2211.08621
(2022)

\bibitem{Rydberg_entanglement_2023}
W. J. Eckner, N. D. Oppong, A. Cao, A. W. Young, W. R. Milner, J. M. Robinson1, J. Ye and A. M. Kaufman,
\textit{Realizing Spin Squeezing with Rydberg Interactions in an Optical Clock},
Nature
(2023)

\bibitem{rosenband_ensemble_2013}
T. Rosenband and D. R. Leibrandt,
\textit{Exponential Scaling of Clock Stability with Atom Number},
arXiv:1303.6357
(2013)

\bibitem{borregaard_ensembles_2013}
J. Borregaard and A. S. Sørensen,
\textit{Efficient Atomic Clocks Operated with Several Atomic Ensembles},
Phys. Rev. Lett.
\textbf{111}, 090802
(2013)

\bibitem{Li_theory_2022}
W. Li, S. Wu, A. Smerzi and L. Pezze,
\textit{Improved Absolute Clock Stability by the Joint Interrogation of Two Atomic Ensembles},
Phys. Rev. A
\textbf{105}, 053116
(2022)

\bibitem{interferometry_2020}
D. Yankelev, C. Avinadav, N. Davidson and O. Firstenberg,
\textit{Atom Interferometry with Thousand-Fold Increase in Dynamic Range},
Sci. Adv.
\textbf{6}, eabd0650
(2020)

\bibitem{Kessler_phase_2014}
E. M. Kessler, P. Kómár, M. Bishof, L. Jiang, A. S. Sørensen, J. Ye and M.D. Lukin,
\textit{Heisenberg-Limited Atom Clocks Based on Entangled Qubits},
Phys. Rev. Lett.
\textbf{112}, 190403
(2014)

\bibitem{Pezze_phase_2020}
L. Pezzè and A. Smerzi,
\textit{Heisenberg-Limited Noisy Atomic Clock Using a Hybrid Coherent and Squeezed State Protocol},
Phys. Rev. Lett.
\textbf{125}, 210503
(2020)

\bibitem{Pezze_phase_2021}
L. Pezzè and A. Smerzi,
\textit{Quantum Phase Estimation Algorithm with Gaussian Spin States},
PRX Quantum
\textbf{2}, 040301
(2021)

\bibitem{zheng_redshift_2023}
X. Zheng, J. Dolde, M.C.~Camrbia, H. M. Lim and S. Kolkowitz,
\textit{A Lab-Based Test of the Gravitational Redshift with a Miniatured Clock Network},
Nat. Commun.
\textbf{14}, 4886 
(2023)

\bibitem{hume_beyond_2016}
D. B. Hume and D. R. Leibrandt,
\textit{Probing Beyond the Laser Coherence Time in Optical Clock Comparisons},
Phys. Rev. A
\textbf{93}, 032138
(2016)

\bibitem{selfcomp0_2011}
St. Falke, H. Schnatz, J.S.R. Vellore Winfred, Th. Middelmann, St. Vogt, S. Weyers, B. Lipphardt, G. Grosche, F. Riehle, U. Sterr and Ch. Lisdat,
\textit{The ${}^{87}$Sr Optical Frequency Standard at PTB},
Metrologia
\textbf{48}, 399--407
(2011)

\bibitem{selfcomp1_2012}
T. L. Nicholson, M. J. Martin, J. R. Williams, B. J. Bloom, M. Bishof, M. D. Swallows, S. L. Campbell and J. Ye,
\textit{Comparison of Two Independent Sr Optical Clocks with $1\times10^{-17}$ Stability at $10^3$ s},
Phys. Rev. Lett.
\textbf{109}, 230801
(2012)

\bibitem{selfcomp2_2015}
A. Al-Masoudi, S. Dörscher, S. Häfner, U. Sterr and Ch. Lisdat,
\textit{Noise and Instability of an Optical Lattice Clock},
Phys. Rev. A
\textbf{92}, 063814
(2015)

\bibitem{ma_phase_1994}
L.-S. Ma, P. Jungner, J. Ye and J. L. Hall,
\textit{Delivering the Same Optical Frequency at Two Places: Accurate Cancellation of Phase Noise Introduced by an Optical Fiber or Other Time-Varying Path},
Optica
\textbf{19}, 1777--1779
(1994)

\bibitem{ptb_phase_2012}
S. Falke, M. Misera, U. Sterr and C. Lisdat,
\textit{Delivering Pulsed and Phase Stable Light to Atoms of An Optical Clock},
Appl. Phys. B.
\textbf{107}, 301--311
(2012)

\bibitem{boyd_nuclear_2007}
M. M. Boyd, T. Zelevinsky, A. D. Ludlow, S. Blatt, T. Zanon-Willette, S. M. Foreman and J. Ye,
\textit{Nuclear Spin Effects in Optical Lattice Clocks},
Phys. Rev. A
\textbf{76}, 022510
(2007)

\bibitem{xia_addressing_2015}
T. Xia, M. Lichtman, K. Maller, A. W. Carr, M. J. Piotrowicz, L. Isenhower and M. Saffman,
\textit{Randomized Benchmarking of Single-Qubit Gates in a 2D Array of Neutral-Atom Qubits},
Phys. Rev. Lett. 
\textbf{114}, 100503
(2015)

\bibitem{trent_addressing_2022}
T. M. Graham, Y. Song, J. Scott, C. Poole, L. Phuttitarn, K. Jooya, P. Eichler, X. Jiang, A. Marra, B. Grinkemeyer, M. Kwon, M. Ebert, J. Cherek, M. T. Lichtman, M. Gillette, J. Gilbert, D. Bowman, T. Ballance, C. Campbell, E. D. Dahl, O. Crawford, N. S. Blunt, B. Rogers, T. Noel and M. Saffman,
\textit{Multi-Qubit Entanglement and Algorithms on a Neutral-Atom Quantum Computer},
Nature
\textbf{604}, 457--462 
2022)


\bibitem{mid_circuit_1}
K. Singh, C. E. Bradley, S. Anand, V. Ramesh, R. White and H. Bernien,
\textit{Mid-Circuit Correction of Correlated Phase Errors Using an Array of Spectator Qubits},
Science
\textbf{380}, 1265--1269
(2023)

\bibitem{mid_circuit_2}
T. M. Graham, L. Phuttitarn, R. Chinnarasu, Y. Song, C. Poole, K. Jooya, J. Scott, A. Scott, P. Eichler and M. Saffman,
\textit{Mid-circuit Measurements on a Neutral Atom Quantum Processor},
Phys. Rev. X
\textbf{13}, 041051
(2023)

\bibitem{mid_circuit_3}
E. Deist, Y.-H. Lu, J. Ho, M. K. Pasha, J. Zeiher, Z. Yan, and D. M. Stamper-Kurn,
\textit{Mid-Circuit Cavity Measurement in a Neutral Atom Array},
Phys. Rev. Lett.
\textbf{129}, 203602
(2022)

\bibitem{mid_circuit_4}
J. W. Lis, A. Senoo, W. F. McGrew, F. Rönchen, A. Jenkins and A. M. Kaufman,
\textit{Midcircuit Operations Using the omg Architecture in Neutral Atom Arrays},
Phys. Rev. X
\textbf{13}, 041035
(2023)

\bibitem{mid_circuit_5}
M. A. Norcia et al.,
\textit{Midcircuit Qubit Measurement and Rearrangement in a 
${}^{171}$Yb Atomic Array},
Phys. Rev. X
\textbf{13}, 041034
(2023)

\bibitem{vector_atomic_2023}
J. D. Roslund, A. Cingöz, W. D. Lunden, G. B. Partridge, A. S. Kowligy, F. Roller, D. B. Sheredy, G. E. Skulason, J. P. Song, J. R. Abo-Shaeer and M. M. Boyd,
\textit{Optical Clocks at Sea}
arXiv:2308.12457
(2023)

\bibitem{clock_gnss_2023}
E. Boldbaatar, D. Grant, S. Choy, S. Zaminpardaz and L. Holden
\textit{Evaluating Optical Clock Performance for GNSS Positioning}
Sensors 
\textbf{2023}, 5998
(2023)

\bibitem{clock_navigation_2021}
T. Schuldt, M. Gohlke, M. Oswald, J. Wüst, T. Blomberg, K. Döringshoff, A. Bawamia, A. Wicht, M. Lezius, K. Voss, M. Krutzik, S. Herrmann, E. Kovalchuk, A. Peters and C. Braxmaier,
\textit{Optical Clock Technologies for Global Navigation Satellite Systems}
GPS Solut
\textbf{25}, 83
(2021)

\bibitem{marti_2018}
G. E. Marti, R. B. Hutson, A. Goban, S. L. Campbell, N. Poli and J. Ye,
\textit{Imaging Optical Frequencies with 100~$\mu$Hz Precision and 1.1~$\mu$m Resolution},
Phys. Rev. Lett.
\textbf{120}, 103201
(2018)

\bibitem{endres_multi_2023}
A. L. Shaw, R. Finkelstein, R. B.-S. Tsai, P. Scholl, T. H. Yoon, J. Choi and M. Endres,
\textit{Multi-ensemble metrology by programming local rotations with atom movements},
arXiv:2303.16885
(2023)

\bibitem{Golay_coils_1958}
M. J. E. Golay,
\textit{Field Homogenizing Coils for Nuclear Spin Resonance Instrumentation},
Rev. Sci. Instrum.
\textbf{29}, 313--315
(1958)

\bibitem{timetrace_pra_2018}
S. de Léséleuc, D. Barredo, V. Lienhard, A. Browaeys and T. Lahaye,
\textit{Analysis of imperfections in the coherent optical excitation of single atoms to Rydberg states},
Phys. Rev. A
\textbf{97}, 053803
(2018)

\bibitem{campbell_2017}
S. L. Campbell, R. B. Hutson, G. E. Marti, A. Goban, N. Darkwah Oppong, R. McNally, L. Sonderhouse, J. M. Robinson, W. Zhang, B. J. Bloom, J. Ye,
\textit{A Fermi-Degenerate Three-Dimensional Optical Lattice Clock},
Science
\textbf{358}, 90--94
(2017)

\bibitem{probe_light_shift}
X. Baillard, M. Fouché, R. Le Targat, P. G. Westergaard, A. Lecallier, Y. Le Coq, G. D. Rovera, S. Bize and P. Lemonde,
\textit{Accuracy Evaluation of an Optical Lattice Clock with Bosonic Atoms},
Opt. Lett.
\textbf{32}, 1812--1814
(2007)

\bibitem{hyper_ramsey}
V. I. Yudin, A. V. Taichenachev, C. W. Oates, Z. W. Barber, N. D. Lemke, A. D. Ludlow, U. Sterr, Ch. Lisdat and F. Riehle,
\textit{Hyper-Ramsey Spectroscopy of Optical Clock Transitions},
Phys. Rev. A 
\textbf{82}, 011804(R)
(2010)

\bibitem{Falke_PTB_2014}
S. Falke, N. Lemke, C. Grebing, B. Lipphardt, S. Weyers, V. Gerginov, N. Huntemann, C. Hagemann, A. Al-Masoudi, S. Häfner, S. Vogt, U. Sterr and C. Lisdat,
\textit{A Strontium Lattice Clock with $3\times10^{-17}$ Inaccuracy and Its Frequency},
New J. Phys.
\textbf{16}, 073023
(2014)

\bibitem{brown_prl_2017}
R.C. Brown, N.B. Phillips, K. Beloy, W.F. McGrew, M. Schioppo, R.J. Fasano, G. Milani, X. Zhang, N. Hinkley, H. Leopardi, T. H. Yoon, D. Nicolodi, T.M. Fortier and A.D. Ludlow,
\textit{Hyperpolarizability and Operational Magic Wavelength in an Optical Lattice Clock},
Phys. Rev. Lett.
\textbf{119}, 253001 
(2017)

\bibitem{katori_lattice_2018}
I. Ushijima, M. Takamoto, and H. Katori, 
\textit{Operational Magic Intensity for Sr Optical Lattice Clocks},
Phys. Rev. Lett.
\textbf{121}, 263202
(2018)

\bibitem{kim_lattice_2023}
K. Kim, A. Aeppli, T. Bothwell and J. Ye,
\textit{Evaluation of Lattice Light Shift at Low 
$10^{-19}$ Uncertainty for a Shallow Lattice Sr Optical Clock},
Phys. Rev. Lett.
\textbf{130}, 113203
(2023)

\bibitem{ptb_latticeLight_2023}
S. Dörscher, J. Klose, S. Maratha Palli and Ch. Lisdat,
\textit{Experimental Determination of the E2-M1 Polarizability of the Strontium Clock Transition},
Phys. Rev. Research
\textbf{5}, L012013 
(2023)

\bibitem{westergaard_tensor_2011}
P.G. Westergaard, J. Lodewyck, L. Lorini, A. Lecallier, E.A. Burt, M. Zawada, J. Millo and P. Lemonde,
\textit{Lattice-Induced Frequency Shifts in Sr Optical Lattice Clocks at the $10^{-17}$ Level},
Phys. Rev. Lett.
\textbf{106}, 210801
(2011)

\bibitem{boyd_thesis_2007}
M. M. Boyd,
\textit{Ph. D. thesis}
(2007).

\bibitem{Runningwave_PRA_2019}
N. Nemitz, A. A. Jørgensen, R. Yanagimoto, F. Bregolin and H. Katori,
\textit{Modeling Light Shifts in Optical Lattice Clocks},
Phys. Rev. A
\textbf{99}, 033424
(2019)

\bibitem{Ushijima_cryogenic_2015}
I. Ushijima, M. Takamoto, M. Dias, T. Ohkubo and H. Katori,
\textit{Cryogenic Optical Lattice Clocks},
Nat. Photon.
\textbf{9}, 185--189
(2015)



\end{thebibliography}

\end{document}